\documentclass[10pt,journal,compsoc]{IEEEtran}
\RequirePackage{etex}

\usepackage{amsmath,amssymb,amsfonts,amsthm}
\usepackage[ruled,vlined,linesnumbered,noend]{algorithm2e}
\usepackage[lambda,advantage,operators,sets,adversary,landau,probability,notions,logic,ff,mm,primitives,events,complexity,asymptotics,keys]{cryptocode}
\usepackage{booktabs}
\usepackage{threeparttable}
\usepackage{enumitem}
\usepackage{stmaryrd} %% for double-struck bracket

\ifCLASSOPTIONcompsoc
  \usepackage{cite}
  \usepackage[caption=false,font=footnotesize,labelfont=sf,textfont=sf]{subfig}
\else
  \usepackage{cite}
  \usepackage[caption=false,font=footnotesize]{subfig}
\fi

\newtheorem{definition}{Definition}
\newtheorem{theorem}{Theorem}

\usepackage{fancyhdr}
\usepackage{hyperref} 
\fancypagestyle{firstpage}{% define a custom header
  \fancyhf{}
  \fancyhead[R]{The paper has been published in IEEE TDSC. For the final version, visit: \href{https://doi.org/10.1109/TDSC.2024.3350206}{https://doi.org/10.1109/TDSC.2024.3350206}}
}

\begin{document}

% \title{CryptoEdge: Practical Secure Aggregation over Encrypted Data at the Edge} 
\title{\textit{TAPFed}: Threshold Secure Aggregation for Privacy-Preserving Federated Learning}

\author{
    Runhua~Xu,~\IEEEmembership{Member,~IEEE}, 
    Bo~Li, 
    Chao~Li, ~\IEEEmembership{Member,~IEEE,}
    James~Joshi,~\IEEEmembership{Fellow,~IEEE}, \\
    Shuai~Ma, ~\IEEEmembership{Senior Member,~IEEE},
    Jianxin~Li, ~\IEEEmembership{Senior Member,~IEEE}

    \IEEEcompsocitemizethanks{
        \IEEEcompsocthanksitem
        Runhua Xu, Bo Li, Jianxin Li and Shuai Ma are affiliated with the School of Computer Science and Engineering at Beihang University, Beijing, China, 100191. 
        Bo Li and Jianxin Li are also associated with Zhongguancun Laboratory in Beijing. 
        Chao Li works at the Beijing Key Laboratory of Security and Privacy in Intelligent Transportation, Beijing Jiaotong University, China, 100044. 
        James Joshi is with the University of Pittsburgh, USA, 15260. \\
        \protect\\
	Emails: runhua@buaa.edu.cn, libo@act.buaa.edu.cn, li.chao@bjtu.edu.cn, jjoshi@pitt.edu, mashuai@buaa.edu.cn, lijx@buaa.edu.cn
    }
}

\IEEEtitleabstractindextext{
\begin{abstract}
    Federated learning is a computing paradigm that enhances privacy by enabling multiple parties to collaboratively train a machine learning model without revealing personal data. However, current research indicates that traditional federated learning platforms are unable to ensure privacy due to privacy leaks caused by the interchange of gradients. To achieve privacy-preserving federated learning, integrating secure aggregation mechanisms is essential. Unfortunately, existing solutions are vulnerable to recently demonstrated inference attacks such as the disaggregation attack.
    This paper proposes \textit{TAPFed}, an approach for achieving privacy-preserving federated learning in the context of multiple decentralized aggregators with malicious actors. \textit{TAPFed} uses a proposed threshold functional encryption scheme and allows for a certain number of malicious aggregators while maintaining security and privacy. We provide formal security and privacy analyses of \textit{TAPFed} and compare it to various baselines through experimental evaluation.
    Our results show that \textit{TAPFed} offers equivalent performance in terms of model quality compared to state-of-the-art approaches while reducing transmission overhead by 29\%-45\% across different model training scenarios. Most importantly, \textit{TAPFed} can defend against recently demonstrated inference attacks caused by curious aggregators,  which the majority of existing approaches are susceptible to.
\end{abstract}

}

\maketitle

\thispagestyle{firstpage}

\section{Introduction}

Federated learning (FL) \cite{konevcny2016federated, mcmahan2017communication} is incredibly promising for collaborative model training amongst several parties - orchestrated by an \textit{aggregator} - without requiring them to provide any of their raw training data, and thus appears to have an elementary privacy guarantee, as it shares only \textit{model updates} such as model weights or gradients, as opposed to the raw private data.
% This approach is compliant with emerging privacy regulations such as the European General Data Protection Regulation (GDPR), the California Consumer Privacy Act (CCPA), and others, and alleviates users' privacy concerns when machine learning (ML) algorithms are applied to their personal data.
Recent studies, however, have shown that private information can still be inferred by exploiting the final ML model, such as through data extraction, membership inference, model inversion, and property inference attacks \cite{shokri2017membership,nasr2019comprehensive,carlini2019secretSharer,ganju2018property}, or by exploiting the exchanged model update during the learning phase \cite{geiping2020inverting,zhu2020deep,jin2020cafe}.
The second category of attack is more FL-specific because it exploits information that is shared during the FL training process, in contrast to the first type of attack, which can be applicable to any ML system and is not just limited to FL systems.
The increasing demand for privacy-preserving FL (PPFL) solutions has resulted in a variety of methods being offered in recent research \cite{xu2019hybridalpha,truex2019hybrid,xu2021fedv,bonawitz2017practical,kadhe2020fastsecagg,so2021turbo,asoodeh2020differentially,zhang2020batchcrypt, xu2022detrust}.
The privacy-preserving aggregation procedure is essential in those PPFL solutions because it can ensure \textit{input privacy} by securing each party's local input model updates and disclosing only the global aggregated model during FL training, thereby effectively preventing or mitigating the second type of attack  \cite{baracaldo2022protecting}.

To achieve privacy-preserving aggregation, various approaches have been proposed, including \textit{privacy-enhancing aggregation} and \textit{secure aggregation}.
Through the use of differential privacy mechanisms, the privacy-enhancing aggregation approaches focus on perturbing model updates with differential privacy (DP) noise \cite{geyer2017differentially,asoodeh2020differentially}, whereas the secure aggregation approaches prevent private information leakage through the use of secure multi-party computation (MPC), pairwise mask technique, and other cryptographic schemes \cite{rathee2023elsa,xu2019hybridalpha,truex2019hybrid,bonawitz2017practical,kadhe2020fastsecagg,so2021turbo,zhang2020batchcrypt}.
In contrast to DP-based privacy-enhancing approaches, secure aggregation techniques ensure the protection of the local model without compromising the aggregated model's accuracy.

\begin{table*}
  \centering
  \caption{Comprehensive Comparison of Various Secure Aggregation Approaches in PPFL}
  \label{tab:summarize}
%   \vspace{-2mm}
  \footnotesize
  \begin{threeparttable}
  \begin{tabular}{lll}
    \toprule
      Representative Approaches & Architecture and Assumption & Limitations \\
    \midrule
      HE \cite{zhang2020batchcrypt,roth2021federated} & Single HbC $A^\dagger$ & Isolation/replay attacks \& single point of failure\\
      Threshold HE \cite{truex2019hybrid} & Single HbC $A^\dagger$  & Stealthy target/gradient inference attacks \& single point of failure\\
      MI-/MC- FE \cite{xu2019hybridalpha, xu2022detrust} & Single HbC $A^\dagger$ & Gradient inference attacks \& single point of failure\\
      (Pairwise) masking \cite{bonawitz2017practical,bell2020secure,roy2022eiffel,ma2023flamingo,bell2023acorn,lycklama2023rofl,kadhe2020fastsecagg,so2021turbo, zheng2022aggregation,liu2022efficient,eltaras2023efficient}  & Single HbC $A^\dagger$ & Gradient inference attacks \& single point of failure\\
      Secure enclave (TEE) \cite{cheng2021separation,zhang2021shufflefl,zhang2023secure}  & Multiple P2P HbC $As^\ddagger$ & Dependency on secure hardware  \\
      MPC(based on secret sharing) \cite{rathee2023elsa} & HbC  $As^\dagger$ (two-server setting) & Scalability, Malicious aggregator \& single point of failure \\
      TAPFed (\textbf{\textit{this work}}) & (threshold) adversarial $As$  & (No need of P2P connection and HbC assumption among $As$)\\
    \bottomrule
  \end{tabular}
  \begin{tablenotes}
    \item[$\dagger$] Single HbC $A$ represents one honest-but-curious aggregator. 
    \item[$\ddagger$] Multiple P2P HbC $As$ denote multiple fully connected peer-to-peer honest-but-curious aggregators.
  \end{tablenotes}
   \end{threeparttable}
   \vspace{-3mm}
\end{table*}

Table \tablename~\ref{tab:summarize} illustrates our examination of the architectures and assumptions of existing secure aggregation techniques, highlighting their potential limitations.
% As illustrated in \tablename~\ref{tab:summarize}, we examine the architectures and assumptions of existing secure aggregation techniques and highlight potential limitations.

To begin with, the majority of known secure aggregation approaches rely on a single centralized aggregator and the assumption of an \textit{honest-but-curious (HbC)} aggregation, leaving the system vulnerable to a variety of attacks. 
For instance, approaches based on threshold homomorphic encryption (THE), multi-input or multi-client functional encryption (MIFE/MCFE), and pairwise masking techniques retain some vulnerability to gradient inference attacks \cite{geiping2020inverting,zhu2020deep,jin2020cafe}.
This is because, even though the aggregation cannot learn each party's input, it can still \textit{access the intermediate aggregated model in plaintext} during multiple rounds of FL training.
On the other hand, while homomorphic encryption (HE) based approaches prevent the curious aggregator from learning the intermediate aggregated models, they leave the aggregator vulnerable to replay attack and isolation attacks \cite{xu2022detrust}.
From the viewpoint of the parties involved, this could lead to a model that doesn't effectively interpret its data or is biased.
A straightforward approach involves spreading trust among multiple decentralized aggregators instead of one centralized aggregator. This results in each aggregator's inability to learn intermediate aggregated models.

Secondly, existing approaches using the two-server setting or multiple-aggregator setting rely on either secure enclave hardware to create a trust execution environment (TEE) or on secure sharing-based multi-party computation techniques to prevent each aggregator from learning the complete intermediate aggregated models \cite{cheng2021separation,zhang2021shufflefl,rathee2023elsa}.
These approaches necessitate complete peer-to-peer collaborative communication between each pair of aggregators, which may pose a scalability problem.
Worse yet, the peer-to-peer collaborative design, unfortunately, heightens the risk of a single point of failure. Should any aggregator fail - whether by accident or intention - it would result in secure aggregation breakdown.

\begin{figure}
    \centering
    \includegraphics[width=.4\textwidth]{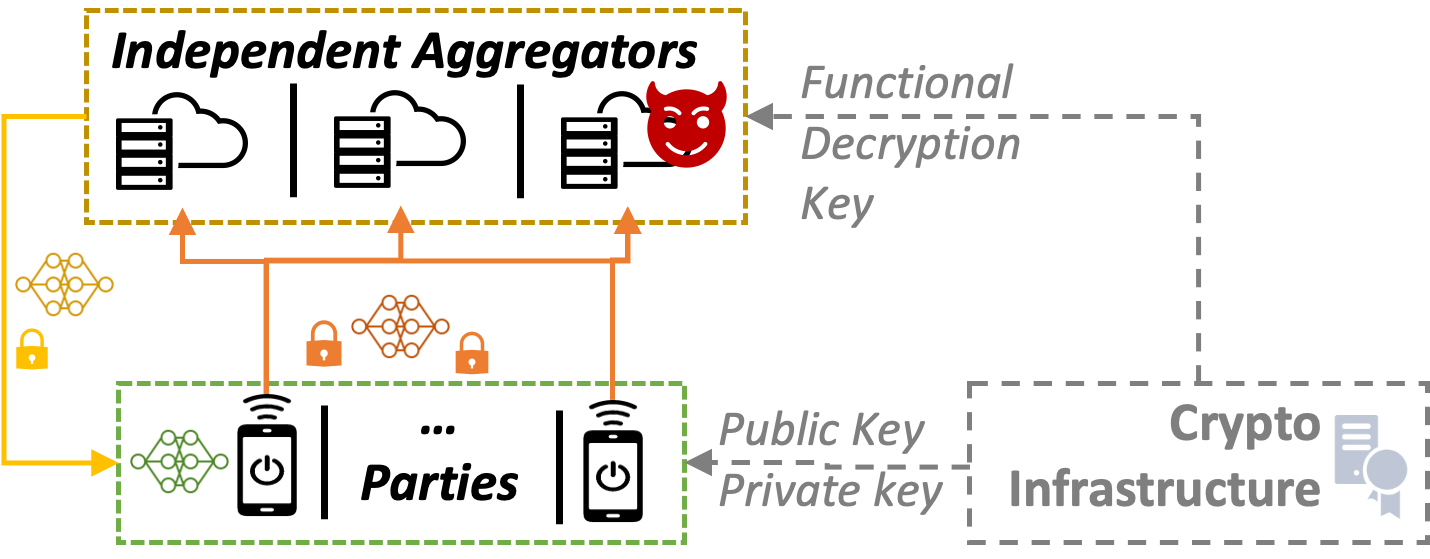}
    \vspace{-2mm}
    \caption{An illustration of \textit{TAPFed} system with three independent aggregators including one malicious aggregator.}
    \label{fig:t-sec-agg}
    \vspace{-5mm}
\end{figure}

To overcome the limitations mentioned above, we introduce a novel framework called \textit{TAPFed} in this paper, which provides \underline{t}hreshold-based secure \underline{a}ggregation for \underline{p}rivacy-preserving \underline{fed}erated learning.

The \textit{TAPFed} employs a multi-aggregator architecture to thwart potential inference attacks, as no single aggregator can access the intermediate model update. It also addresses the single point of failure issue by eliminating the need for fully connected peer-to-peer communication among aggregators, a dependency in existing frameworks \cite{cheng2021separation,zhang2021shufflefl,rathee2023elsa}. To accomplish this, we introduce an innovative \textit{threshold functional encryption (TFE)} scheme that enables secure aggregation with independent and decentralized aggregators while tolerating a limited number of malicious ones. Importantly, each aggregator remains oblivious to others' existence and cannot learn the intermediate aggregated model.

The \textit{TAPFed} framework, depicted in Figure~\ref{fig:t-sec-agg}, consists of multiple parties and independent aggregators. In each round of FL training within \textit{TAPFed}, every party sends encrypted local model updates to all connectable aggregators who then perform secure aggregation using the provided functional decryption key. 
Subsequently, each aggregator delivers the intermediate aggregated model (still encrypted) back to each party. 
The parties can then obtain the aggregated global model via a threshold secure aggregation mechanism for use in the next FL training round. 
It's important to note that in \textit{TAPFed}, all aggregators work independently without needing peer-to-peer communication with others - a unique feature that reduces single point of failure risks if an aggregator malfunctions.
Note that \textit{TAPFed} primarily concentrates on cross-silo federated learning, aligning with the current secure aggregation research that utilizes computational cryptographic solutions.

The key \textit{\textbf{contributions}} of this paper are as follows:

\noindent\textbf{Threshold Functional Encryption}: 
We propose a \textit{threshold functional encryption (TFE)} scheme for securely computing the inner product that is based on the \textit{Decisional Diffie-Hellman (DDH)} assumption in the integer group with \textit{selective simulation-based (SEL-SIM)} security.
Taking into account the computational capability of the party, which may include IoTs, we construct the TFE scheme using only the basic DDH assumption rather than more complex and computationally inefficient security assumptions like pairing-based security in bilinear groups and matrix decision-based DDH assumptions.
% The proposed \textit{TAPFed} framework is built on the TFE cryptosystem.
% TODO: first tmcfe scheme definition and constructions. add prove what kind of security proof.

\noindent\textbf{Threshold Secure Aggregation for Privacy-Preserving Federated Learning}: 
In addition, we present a privacy-preserving federated learning (PPFL) framework, termed \textit{TAPFed}, which supports threshold secure aggregation based on the TFE cryptosystem outlined previously.
\textit{TAPFed} can effectively defend against recently demonstrated inference attacks and permits a limited number of adversarial aggregators.
Both the classic average fusion method and the weighted fusion method are supported by \textit{TAPFed}.
Without the need for secure enclave hardware or the peer-to-peer necessity for decentralized aggregators, \textit{TAPFed} can withstand isolation and gradient inference attacks that the majority of previous approaches are vulnerable to.

\noindent\textbf{Security and Privacy Analysis}:
In addition, we provide formal security proof for the proposed TFE scheme and a thorough analysis of the \textit{TAPFed} framework's privacy implications. 
Our proof and analysis demonstrate that \textit{TAPFed} meets its intended security requirements and can successfully defend against various recently demonstrated attacks.

\noindent\textbf{Experimental Evaluation}: 
In this paper, we performed two primary evaluations. 
Firstly, we implemented the \textit{TAPFed} framework and compared it experimentally to a variety of baseline solutions. 
Secondly, we evaluated the performance of \textit{TAPFed} under different settings, such as varying sizes of involved parties and length of encoding precision.
In summary, our evaluation demonstrates that the \textit{TAPFed} framework outperforms existing techniques in terms of system performance while also providing improved security and privacy guarantees.

% \vspace{0.1in}
% \noindent\textbf{Organization}. {\color{blue}(TO BE UPDATED)}
% The rest of the paper is organized as follows.
% In Section~\ref{sec:pre}, we introduce the necessary background and preliminaries.
% We present our proposed TFE scheme in Section~\ref{sec:tfe} and the proposed \textit{CryptoEdge} framework in Section~\ref{sec:ce}.
% The security and privacy analyses are presented in Section~\ref{sec:sp}, and the experimental evaluation is presented in Section~\ref{sec:eval}.
% We discuss related work in Section~\ref{sec:related} and conclude the paper in Section~\ref{sec:conclusion}.

% \vspace{-3mm}
\section{Related Work}
\label{sec:related}

\subsection{Privacy-Preserving Federated Learning}

The concept of federated learning (FL), initially presented in \cite{mcmahan2017communication,konevcny2016federated}, is a distributed machine learning framework that aims to collaboratively train a machine learning model through exchanging model updates such as weights or gradients, rather than sharing personal data. 
While the FL paradigm appears to offer privacy assurances at first glance, the exchange of model updates still poses potential risks. Curious aggregators could potentially extract private information from the shared model updates, as demonstrated in \cite{shokri2017membership,nasr2019comprehensive,zhu2020deep}, rendering the privacy guarantees insufficient.
To prevent leakage of private information from parties' output, \textit{secure aggregation} plays a crucial role in FL to ensure that the central aggregator only obtains the fused model updates without learning any input model updates from parties.

Note that it's important to distinguish between privacy issues and security issues in the context of federated learning. 
This paper won't discuss attacks like poisoning, backdoor or robustness-related issues, as summarized in \cite{jere2020taxonomy,zhang2023secure}, which aim to disrupt the functionality or integrity of the federated learning system.
In the context of privacy protection in federated learning, from the perspective of aggregator settings, secure aggregation can be divided into two categories: single honest-but-curious aggregator and multi-aggregator settings.

\subsection{Secure Aggregation in Single Honest-but-Curious Aggregator Setting}

In the single honest-but-curious aggregator setting, the aggregator is assumed to be honest-but-curious, which means that it will follow the protocol but may attempt to learn private information from the obtained intermediate aggregated model updates.
The majority of existing secure aggregation approaches are based on this setting, where the secure aggregation solutions employ a variety of techniques:(\romannumeral1) approaches based on secure multi-party computation protocols using garbled circuits \cite{pinkas2009secure,mohassel2015fast,wang2017global}; 
(\romannumeral2) approaches based on anonymous communication using mix-nets \cite{chaum1981untraceable} or DC-nets\cite{chaum1988dining}; 
(\romannumeral3) approaches based on cryptosystems such as partially/fully HE or FE \cite{xu2019hybridalpha,truex2019hybrid,zhang2020batchcrypt}; 
(\romannumeral4) approaches based on pairwise masking \cite{bonawitz2017practical,bell2020secure,lycklama2023rofl,bell2023acorn,ma2023flamingo,roy2022eiffel,liu2022efficient,eltaras2023efficient}.

The methods described in \cite{pinkas2009secure,mohassel2015fast,wang2017global} rely on secure multi-party computing protocols that use garbled circuit techniques. 
However, these approaches have a significant drawback in that they require the exchange of large garbled tables for each circuit gate, leading to a communication overhead. 
Alternatively, anonymous communication techniques like DC-nets \cite{chaum1988dining} or mix-nets \cite{chaum1981untraceable} offer a different approach by shielding the connections between the privacy-sensitive data and the parties, instead of solely preventing the disclosure of private information.

Most of the emerging and promising secure aggregation approaches in the single honest-but-curious aggregator setting rely on advanced cryptographic systems, such as partially or fully homomorphic encryption (HE) and multi-input or multi-client functional encryption (FE) \cite{xu2019hybridalpha, truex2019hybrid,zhang2020batchcrypt}, TEE-based approach \cite{zhang2023secure}, and lightweight cryptographic primitives such as pairwise masking \cite{bonawitz2017practical,bell2020secure,lycklama2023rofl,bell2023acorn,ma2023flamingo,roy2022eiffel,zheng2022aggregation}.

Pairwise masking-based methods can accommodate a large number of parties with consideration for dropouts, but their design requires multiple rounds of communication between the central aggregator and parties for a single round of secure aggregation.
Trust execution environment (TEE) based approaches, also known as confidential computing-based approaches, rely on specialized hardware that allows for secure enclaves, such as Intel SGX, AMD PSP, and ARM TrustZone. However, these secure enclave hardware are not widely available and still have limited memory space for secure processing.
Rather than following the path of pairwise masking, which may result in increased communication overhead, this paper concentrates on the use of advanced cryptographic techniques, which represent a viable emerging trend for resolving secure aggregation problems in a simple communication topology without the need for special hardware support.

Instead of relying on fully homomorphic encryption (HE) based solutions that have computational limitations resulting in not practical for large-scale secure aggregation of model updates and suffer from potential isolation or replay attacks, alternative solutions such as multi-input functional encryption (MIFE) or multi-client functional encryption (MCFE) demonstrate promise in terms of both computation and communication efficiency \cite{xu2019hybridalpha,ryffel2019partially}.
However, MIFE/MCFE-based solutions still suffer from recently demonstrated model disaggregation attacks that are launched by the central curious aggregator by exploiting the information of aggregated model and fusion weights from multiple FL training rounds.

\subsection{Secure Aggregation in Multi-Aggregator Setting}

To address potential privacy breaches from an honest-but-curious aggregator, a simple solution is to distribute trust among several decentralized aggregators rather than relying on one centralized source. This prevents any single aggregator from learning intermediate aggregated models, thereby eliminating the possibility of inference attacks.

Existing secure aggregation approaches in multi-aggregator setting lies in two categories: (\romannumeral1) approaches based on secure multi-party computation protocols using secret sharing primitives \cite{rathee2023elsa} and 
(\romannumeral2) secure enclave \cite{cheng2021separation,zhang2021shufflefl}. 

The approaches based on secure multi-party computation protocols using secret sharing primitives \cite{rathee2023elsa} are designed for the two-server setting, which requires a peer-to-peer connection to exchange shared information, resulting in a complex communication topology.
The same limitation also occurs at the approaches based on secure enclave-based secret sharing approaches \cite{cheng2021separation,zhang2021shufflefl}. 
Moreover, the two-server setting is vulnerable to a single point of failure, as well as the feasibility and likelihood of collusion among aggregators.

To address the above-mentioned secure aggregation challenges, we focus on the path of MCFE techniques to take advantage of computation and communication efficiency and solve the disaggregation problem with the setting of untrusted multiple aggregators by proposing a threat variant of functional encryption primitives to achieve the threat model in the FL environment.

Note that our study, which follows a similar methodology as described in \cite{xu2019hybridalpha, truex2019hybrid}, is also compatible with differential privacy mechanisms. These are another fundamental technique to achieve privacy-preserving federated learning by adding noise to perturb the model and provide an output privacy guarantee for the final trained model \cite{wei2020federated}.

% \vspace{-3mm}
\section{Threshold Functional Encryption}
\label{sec:tfe}

\subsection{Motivation of Threshold Functional Encryption}
\label{sec:tfe:overview}

As we've previously discussed, \textit{TAPFed} is constructed on a multi-aggregator architecture to ward off potential inference attacks. To address the issues of single-point failure and scalability, a new secure aggregator protocol should be developed based on an innovative cryptographic primitive. This differs from existing secret sharing-based multi-party computation primitives as it doesn't necessitate any peer-to-peer communication between aggregators.

To accomplish this, we propose a groundbreaking threshold functional encryption (TFE) scheme. This allows for secure aggregation with independent and decentralized aggregators without any fully connected peer-to-peer communication, while tolerating a limited number of malicious ones.

Given that threshold functional encryption aligns with computational cryptographic primitives, it's reasonable to question why we wouldn't leverage existing Homomorphic Encryption (HE), threshold HE, or Functional Encryption (FE) schemes. After all, these are viable alternatives and promising strategies in the field of secure aggregation \cite{xu2020revisiting}.
We do not employ these alternative techniques directly for the following reasons:
(\romannumeral1) HE- or FE-based techniques rely on centralized aggregator settings, resulting in isolation and disaggregation attacks, as demonstrated in recent literature \cite{xu2022detrust};
(\romannumeral2) As the most relevant work, threshold HE (e.g., threshold Paillier cryptosystem) is inefficient for handling complex applications over encrypted data, and it also suffers from the same disaggregation problem as FE-based techniques.

Recent multi-input FE (MIFE) or multi-client FE (MCFE) systems \cite{abdalla2018multi,chotard2018decentralized,abdalla2019single} have shown the applicability of the computation-efficient DDH assumption in FE cryptosystems, making it more suitable for IoT devices than pairing and garbled circuit-based alternatives.
Considering their promising use in the creation of practical PPML applications \cite{xu2019cryptonn,xu2019hybridalpha,xu2021nn,xu2021fedv}, we propose a new threshold MCFE scheme based on the DDH assumption that supports secure aggregation in decentralized aggregators settings and handling issues as previously analyzed.

In short, \textit{TAPFed} leverages TFE cryptographic primitive for its dual benefits: it uses a multi-aggregator structure that naturally wards off recently demonstrated inference attacks that require to access intermediate global model update and employs a simple communication topology free from burdensome peer-to-peer exchanges among nodes.
Crucially, our secure aggregation protocol, based on threshold functional encryption, can withstand collusion from a limited number of malicious aggregators - a feature not offered by existing multi-aggregator solutions. This advantage is ensured by the design of the underlying cryptographic primitive's threshold functionality.

\subsection{Preliminaries: Functional Encryption Definitions}
\label{sec:pre:fe}
Functional encryption (FE) is a group of cryptosystems that allow functions to be computed over encrypted data, with the resulting function value being in plaintext.
Following the initial definition from \cite{boneh2011functional} and \cite{abdalla2018multi}, we present the notion of functionality, functional encryption scheme,  security assumption and security definition.

\begin{definition}[Functionality \cite{boneh2011functional}]
A \textit{functionality} $\mathcal{F}$ defined over $(K,X)$ is a function $\mathcal{F}:K \times X \to \Sigma \cup \{\perp\}$ where $K$ is the \textit{key space}, $X$ is the \textit{message space} and $\Sigma$ is the \textit{output space} and $\perp$ is a special string not contained in $\Sigma$. 
\end{definition}
\begin{definition}[Functional Encryption Scheme \cite{boneh2011functional}]
A \textit{functional encryption (FE)} scheme for functionality $\mathcal{F}$ is a tuple $\mathcal{E}_\text{FE}$ = ($\mathtt{Setup}, \mathtt{KeyDerive}, \mathtt{Encrypt}, \mathtt{Decrypt}$) of four algorithms:
\begin{itemize}
    \item $\mathtt{Setup}(1^\lambda)$ outputs \textit{public} and \textit{master secret keys ($\pk_{\text{m}}, \sk_{\text{m}}$)} for \textit{security parameter} $\lambda$;
    \item $\mathtt{KeyDerive}(\sk_{\text{m}}, k)$ outputs \textit{secret key} $sk_{k}$ given an input a master secret key, $\sk_{\text{m}}$, and a \textit{key}, $k \in K$;
    \item $\mathtt{Encrypt}(\pk_{\text{m}}, x)$ outputs ciphertext $ct$ given an input a public key, $\pk_{\text{m}}$, and a \textit{message}, $x \in X$;
    \item $\mathtt{Decrypt}(\pk_{\text{m}}, ct, sk_x)$ outputs $z \in \Sigma \cup\{\perp\}$.
\end{itemize}
\end{definition}
% Note that \textit{functionality} is undeﬁned when the key is not in the keyspace, or the message is not in the message space.
In addition, the \textit{correctness} of $\mathcal{E}_\text{FE}$ is described as $\forall (\pk_{\text{m}}, \sk_{\text{m}}) \gets \mathtt{Setup}(1^\lambda)$, $\forall k \in K, x \in X$, for $\sk_k \gets \mathtt{KeyDerive}(\sk_{\text{m}}, k)$ and $ct \gets \mathtt{Encrypt}(\pk_{\text{m}}, x)$, we have $\mathtt{Decrypt}(\pk_{\text{m}}, ct, \sk_x) = \mathcal{F}(x,k)$  whenever $\mathcal{F}(x,k) \ne \perp$, except with negligible probability.

% \subsubsection{Security of Functional Encryption}

\begin{definition}[Selective Simulation-based Secure FE \cite{abdalla2018multi}]
A functional encryption $\mathcal{E}_{\text{FE}}$ for functionality $\mathcal{F}$ is selective simulation-based secure (SEL-SIM-secure) if there exist \textit{PPT} simulator algorithms $\mathcal{E}^{\text{SIM}}_{\text{FE}}$=$(\mathtt{Setup}^{\text{SIM}}$, $\mathtt{KeyDerive}^{\text{SIM}}$, $\mathtt{Encrypt}^{\text{SIM}}$, $\mathtt{Decrypt}^{\text{SIM}})$ such that for every stateful PPT adversary $\mathcal{A}$ and $\lambda \in \NN$, the following two distributions are computationally indistinguishable:

\begin{minipage}{.22\textwidth}
\procedure[codesize=\footnotesize]{$ \mathtt{Exp}\;  \textbf{REAL}^{\mathcal{E}_{\text{FE}}}_{\text{SEL}}(1^{\lambda}, \mathcal{A})$}{%
\{x_i\}_{i\in[n]} \gets \mathcal{A}(1^{\lambda}, \mathcal{F}) \\
(\pk_{\text{m}},\sk_{\text{m}}) \gets \mathtt{Setup}(1^{\lambda}, \mathcal{F}) \\
\forall i\in[n], ct_i \gets \mathtt{Enc}_{\pk_{\text{m}}}(i, x_i) \\
\alpha \gets \mathcal{A}^{\mathtt{KeyDer}(\sk_{\text{m}})}_{\pk_{\text{m}}}(\{ct_i\}_{i\in[n]}) \\
\textbf{Output}: \alpha \\[1mm][\hline]
}
\end{minipage}
\begin{minipage}{.22\textwidth}
\procedure[codesize=\footnotesize]{$ \mathtt{Exp}\; \textbf{IDEAL}^{\mathcal{E}_{\text{FE}}}_{\text{SEL}}(1^{\lambda}, \mathcal{A})$}{%
\{x_i\}_{i\in[n]} \gets \mathcal{A}(1^{\lambda}, \mathcal{F}) \\
(\pk_{\text{m}}^{\text{SIM}},\sk_{\text{m}}^{\text{SIM}}) \gets \mathtt{Setup}^{\text{SIM}}(1^{\lambda}, \mathcal{F}) \\
\forall i\in[n], ct_i \gets \mathtt{Enc}^{\text{SIM}}_{\pk_{\text{m}}^{\text{SIM}}}(i) \\
\alpha \gets \mathcal{A}^{\mathcal{O}(\cdot)}_{\pk_{\text{m}}^{\text{SIM}}}(\{ct_i\}_{i\in[n]}) \\
\textbf{Output}: \alpha \\[1mm][\hline]
}
\end{minipage}

The oracle $\mathcal{O}(\cdot)$ in the ideal experiment above is given access to another oracle that, given $f\in\mathcal{F}$, returns $f(x_1, ..., x_n)$, and then $\mathcal{O}(\cdot)$ returns $\mathtt{KeyDer}^{\text{SIM}}(\sk_{\text{m}}^{\text{SIM}}, f, f(x_1, ..., x_n))$. 
\end{definition}

Note that for every stateful adversary $\mathcal{A}$, we define its advantage as $\advantage{\text{SEL-SIM}}{\adv,\mathcal{E}_\text{FE}} = |\Pr[\textbf{REAL}^{\mathcal{E}_\text{FE}}_\text{SEL}(1^{\lambda}, \mathcal{A})=1]-\Pr[\textbf{IDEAL}^{\mathcal{E}_\text{FE}}_\text{SEL}(1^{\lambda}, \mathcal{A})]|$ and we require that for every PPT $\adv$, there exists a negligible function $\negl$ such that $\forall \lambda\in\NN, \advantage{\text{SEL-SIM}}{\adv,\mathcal{E}_\text{FE}} = \negl$.

% \subsubsection{The Decisional Diffie-Hellman Assumption}
\noindent\textbf{Decisional Diffie-Hellman (DDH) Assumption}.
The DDH assumption states that the tuples $(g, g^a, g^b, g^{ab})$ and $(g, g^a, g^b, g^c)$ are computationally indistinguishable, where $a,b,c \in \ZZ_p$ are chosen independently and uniformly at random.

% \vspace{-3mm}
\subsection{Definition of Threshold Functional Encryption}
\label{sec:tfe:definition}

\noindent\textbf{Notation}.
As a brief notational introduction to the following threshold FE (TFE) presentation,  
let $\mathtt{GroupGen}(1^{\lambda})$ be a probabilistic polynomial-time algorithm that takes as input a security parameter $1^\lambda$, and outputs a triplet $(\GG, p, g)$, where $\GG$ is a group of order $p$ that is generated by $g \in \GG$, and $p$ is a $\lambda$-bit prime number.
Furthermore, let $r \sample \ZZ_p$ denote the assignment to $r$ an element chosen uniformly at random from integer group $\ZZ_p$.
We use $\llbracket x \rrbracket$ to denote encrypted $x$. 
A lowercase bold variable such as $\pmb{\alpha}^{1\times \eta}$ represents a vector with length, $\eta$.
A capital bold variable such as $\pmb{W}^{n\times\eta}$ denotes a matrix with $n$ rows and $\eta$ columns.

We define \textit{threshold functional encryption (TFE) for functionality $\mathcal{F}$} scheme as follows.
\begin{definition}[Threshold Functional Encryption Scheme (TFE)] 
A \textit{t-of-s threshold functional encryption} for functionality $\mathcal{F}$ is a tuple of following six algorithms:
\begin{itemize}[leftmargin=*]
    \item[-] $\mathtt{Setup(}\lambda\textit{)}$ outputs public parameter and master secret key, ($\mathsf{pp},\mathsf{msk}$), based on security parameter, $\lambda$.
    
    \item[-] $\mathtt{SKDistribute(}\mathsf{pp},\mathsf{msk}, e_{\text{idx}}\textit{)}$ distributes secret key, $\sk_{\text{idx}}$, for encryption entity $e_{\text{idx}}$ on input master keys ($\pk_{\text{m}}, \sk_{\text{m}}$).
    
    \item[-] $\mathtt{DKGenerate(}\mathsf{pp},\mathsf{msk}, \mathcal{K}, d_{\text{idx}}, \mathcal{L} \textit{)}$ generate functional decryption key, $\mathsf{dk}_{\text{idx}}$, for a decryption entity, $d_{\text{idx}}$ on input master keys, ($\pk_{\text{m}}, \sk_{\text{m}}$), and a label $\mathcal{L}$, and a vector from $\mathcal{K}$.
    
    \item[-] $\mathtt{Encrypt(}\sk_{\text{idx}}, \mathcal{X}, \mathcal{L} \textit{)}$ outputs ciphertext of $\llbracket\mathcal{X}\rrbracket$ on input vector from $\mathcal{X}$, secret key $\sk_{\text{idx}}$ and a label $\mathcal{L}$. 
    
    \item[-] $\mathtt{ShareDecrypt(}\mathsf{pp}, ct, \mathcal{K}, \mathsf{dk}_{\text{idx}}, \mathcal{L}, S\textit{)}$ outputs a partially decrypted ciphertext, $\llbracket\mathcal{X}\rrbracket^{'}$, on input a ciphertext, $ct$, a public parameter, $\mathsf{pp}$, a vector from $\mathcal{K}$, a label $\mathcal{L}$, and a functional private key, $\mathsf{dk}_{\text{idx}}$, a selected sub-set of decryption parties $S$.
    
    \item[-] $\mathtt{CombineDecrypt(}\mathsf{pp}, ct^{'}, \mathcal{L}\textit{)}$ outputs functionality result on input public parameter, $\mathsf{pp}$, a label $\mathcal{L}$, and partially decrypted ciphertext, $\llbracket\mathcal{X}\rrbracket^{'}$.
\end{itemize}
\label{def:th-feip}
\end{definition}

% \vspace{-3mm}
\subsection{Proposed Threshold Multi-Client FE Scheme}
\label{sec:tfe:miip}

% {\color{blue} TODO: upgrade the construction from current Threshold MIFE to threshold MCFE. $\to$ DONE}

% Here, we present our proposed \textit{t-of-s threshold functional encryption} scheme for functionality $\mathcal{F}_{\text{MIIP}}$, whose security is based on the plain \textit{Decisional Diffie-Hellman} (DDH) assumption, where $t$ of $s$ decryption parties are allowed to collaboratively acquire a inner-product over multiple encrypted vectors.

\noindent\textbf{Functionality of $\mathcal{F}_{\text{MCIP}}$}.
In this paper, we mainly focus on the \textit{inner-product functionality} over the integers.
Let $\mathcal{F}_{\text{IP}}$ be a family of \textit{inner-product functionality} with message space $\mathcal{X}$ and key space $\mathcal{K}$ both consisting of vectors in $\ZZ^\eta_p$ of norm bounded by $p$ of length $\eta$.
Here, we focus on \textit{multiple clients} inner-product $\mathcal{F}_{\text{MIIP}}$, defined as follows:
\begin{align*}
    f_{\text{MCIP}}(\{(\pmb{x}_{i}, l_{\pmb{x}_i})\},\{\pmb{y}, l_{\pmb{y}}\}) = \sum_{i\in[n]}\sum_{j\in[\eta_i]}(x_{ij}y_{\sum^{i-1}_{k=1}\eta_{k}+j}) \\
    \text{s.t.}\;|\pmb{x}_i|=\eta_{i}, |\pmb{y}| = \sum_{i\in[1,...,n]}\eta_{i}, \forall i\in[1,...,n]:l_{\pmb{x}_i} = l_{\pmb{y}}
\end{align*}
where $f_{\text{MCIP}} \in \mathcal{F}_{\text{IP}}$, $\pmb{x}_i \in \mathcal{X}$, $\pmb{y} \in \mathcal{K}$, and $l \in \mathcal{L}$.
Also, the total length of $\pmb{x}_i$ should be equal to the length of vector $\pmb{y}$.

\noindent\textbf{Construction}.
Beginning with the multi-client FE (MCFE) scheme, our threshold MCFE (tMCFE) scheme for $\mathcal{F}_{\text{MCIP}}$ is constructed as follows:

\begin{itemize}[leftmargin=*]
    \item $\mathtt{Setup(}\lambda, \eta, t, s, n\textit{)}$:
    The algorithm first generates a triplet from the integer group, as $(\GG, p, g)$ $\sample$ $\mathtt{GroupGen}(1^{\lambda})$, on given security parameter $\lambda$ as input and defines a full-domain hash function $\mathsf{H}$ onto $\GG$.
    Then, it randomizes a matrix of $1 \times \eta$ samples and two matrix samples with size $n \times \eta$, represented as:
    $
    \pmb{\alpha}^{1 \times \eta} \sample \ZZ^\eta_p, 
    \pmb{W}^{n \times \eta} \sample \ZZ^\eta_p, 
    \pmb{U}^{n \times \eta} \sample \ZZ^\eta_p.
    $
    The public parameter $\mathsf{pp}$ and master private key $\mathsf{msk}$ are defined as follows: $\mathsf{pp} = (\GG, p, g, t, s, n, \mathsf{H})$, $\mathsf{msk}= (\pmb{W}^{n \times \eta}, \pmb{U}^{n \times \eta}, g^{\pmb{\alpha}}, \{g^{\pmb{\alpha}^{\intercal}\pmb{W}_i}\}_{i \in \{1,...,n\}})$.
    % \begin{align*}
    %     \mathsf{pp} &= (\GG, p, g, t, s, n, \mathsf{H}) \\
    %     \mathsf{msk} &= (\pmb{W}^{n \times \eta}, \pmb{U}^{n \times \eta}, g^{\pmb{\alpha}}, \{g^{\pmb{\alpha}^{\intercal}\pmb{W}_i}\}_{i \in \{1,...,n\}}) \\
    % \end{align*}

    \item $\mathtt{SKDistribute(}\mathsf{pp},\mathsf{msk}, e_{i}\textit{)}$:
    Given master keys, for encryption entity $e_{\text{idx}} \in \{1,...,n\}$, the algorithm distributes the secret keys as 
    $\sk_{e_{\text{idx}}} = (\mathsf{pp},  g^{\pmb{\alpha}}, g^{\pmb{\alpha}^{\intercal}\pmb{W}_{e_{\text{idx}}}}, \pmb{U}_{e_{\text{idx}}}).$
    
    \item $\mathtt{DKGenerate(}\mathsf{pp},\mathsf{msk}, \pmb{y}, d_{\text{idx}}, l\textit{)}$:
    The algorithm takes master keys, functionality-related vector $\pmb{y} =(\pmb{y}_1, \pmb{y}_2, ..., \pmb{y}_n)$, and distributes functional decryption key for corresponding decryption entity $d_{\text{idx}}$ and label $l$.
    The algorithm first defines a set of polynomial functions $(f^{(0)}(x)=\sum^{t-1}_{k=0}a_{k}x^{k}, \{f^{(i)}(x)= \sum^{t-1}_{k=0}b_{i,k}x^k\}_{i\in\{1,...,n\}})$, 
    where $a_{k} \sample \ZZ_p$, $b_{i,k} \sample \ZZ_p$, $a_0 = \mathsf{H}(l)\sum^{n}_{i=1}\langle\pmb{y}_i,\pmb{U}_i\rangle$, and $b_{i,0} =\langle\pmb{y}_i,\pmb{W}_i\rangle$.
    The algorithm then generates a set of functional decryption keys $\mathsf{dk} = \{v_{j,0}, v_{j,1}\}_{j\in\{1,...,s\}}$, where $v_{j,0} = f^{(0)}(j), v_{j,1} = \{f^{(i)}(j)\}_{i\in\{1,...,n\}}$.
    For the partial decryption entity $d_{\text{idx}}\in\{1,...,s\}$, the algorithm distributes the functional private key $\mathsf{dk}_{d_\text{idx}} = (\mathsf{pp}, v_{\text{idx}, 0}, v_{\text{idx}, 1})$.
    
    \item $\mathtt{Encrypt(}\sk_{i},\pmb{x}_{i}, l\textit{)}$: 
    For the encryption entity $e_{\text{idx}} \in\{1,...,n\}$, the algorithm takes as input $\sk_{i}$ and $\pmb{x}_{i}$ with specified label $l$, and returns ciphertext  $\llbracket\pmb{x}_{i}\rrbracket$.
    It first chooses a random element $r_i \sample \ZZ_p$ and computes the ciphertext $\llbracket\pmb{x}_{i}\rrbracket = (ct_{i,0}, ct_{i,1})$ as follows:
    $$
    ct_{i,0} = g^{\pmb{x}_{i}+\mathsf{H}(l)\pmb{U}_{i}} \circ (g^{\pmb{\alpha}^\intercal\pmb{W}_{i}})^{r_{i}},
    ct_{i,1} 
        = \prod(g^{\pmb{\alpha}})^{r_{i}}.
    $$
    % \begin{align*}
    %     ct_{i,0} &= g^{\pmb{x}_{i}+\mathsf{H}(l)\pmb{U}_{i}} \circ (g^{\pmb{\alpha}^\intercal\pmb{W}_{i}})^{r_{i}},\\
    %     ct_{i,1} 
    %     &= \prod(g^{\pmb{\alpha}})^{r_{i}} 
    %     = \prod_{k \in\{1,...,\eta\}}g^{\alpha_{k}r_{i}}.
    % \end{align*}
    Note that the symbol $\circ$ denotes the element-wise multiplication. For instance, $\pmb{x}^{1\times\eta}\circ\pmb{Y}^{n\times\eta} \to \pmb{Z}^{n\times\eta}$ denotes the element-wise multiplication of elements at the corresponding position in $\pmb{x}$ and each row of $\pmb{Y}$.
    
    \item $\mathtt{ShareDecrypt(}\mathsf{pp}, \{\llbracket\pmb{x}_{i}\rrbracket\}_{i\in\{1,...,n\}}, \pmb{y}, \mathsf{dk}_{j}, S\textit{)}$:
    The algorithm takes the ciphertext $\{\llbracket\pmb{x}_{i}\rrbracket\}_{i\in\{1,...,n\}}$, the public parameter $\mathsf{pp}$ and vector $\pmb{y} = \{\pmb{y}_i\}_{i\in\{1,...,n\}}$ associated decryption key $\mathsf{dk}_{j}$ from an authorized sub-set $S$, where $|S| \ge t$.
    For the sharing decryption entity $d_{j}\in S$ with $\mathsf{dk}_{j}$, it outputs the partially decrypted ciphertext $\llbracket\pmb{ct}\rrbracket^{'}_{j} = (ct^{'}_{j,0}, ct^{'}_{j,1}, ct^{'}_{j,2})$ as follows:
    \begin{align*}
        ct^{'}_{j,0} &= \prod_{i\in\{1,...,n\}}ct_{i,0}^{\circ\pmb{y}_i},\\
        ct^{'}_{j,1} &= \{(ct_{i,1})^{v_{j,1}L_{j}(j)}\}_{i\in\{1,...,n\}},\\
        ct^{'}_{j,2} &= g^{v_{j,0}L_{j}(j)},
    \end{align*}
    where $L_j(j)$ is the Lagrange basis polynomials defined as $\prod_{j^{'}\in S, j^{'} \neq j}\frac{-j^{'}}{j - j^{'}}$.
    Note that $\pmb{X}^{(n\times m)\circ\pmb{y}^{(1\times m)}} \to \pmb{Z}^{n\times m}$ represents element-wise exponentiation of elements at the corresponding position in $\pmb{X}$ and $\pmb{y}$.
    
    \item $\mathtt{CombineDecrypt(}\mathsf{pp}, \{\llbracket\pmb{ct}\rrbracket^{'}_{j}\}_{j\in\{1,..,s'\}}\textit{)}$: 
    The algorithm takes all received ciphertext $\{\llbracket\pmb{ct}\rrbracket^{'}_{j}\}_{j\in\{1,..,s'\}}$ and returns the inner-product $\langle\{\pmb{x}_i\}_{i\in\{1,...,n\}},\pmb{y}\rangle$.
    $\forall ct^{'}_{j,0} \in \{[ct]^{'}_{j}\}_{j\in\{1,..,s'\}}$, the algorithm verifies they are all equal. 
    If the \textit{\textbf{verification}} is not passed, it returns the stop symbol; otherwise, let $C=ct^{'}_{1,0}$.
    Then, the algorithm returns the final combined decryption results as follows:
    $$
    D = \frac{C}{\prod_{i}\prod_{j}ct^{'}_{i,j,1} \cdot \prod_{j}(ct^{'}_{j,2})^2},
    $$
    where $i\in\{1,...,n\}$ and $j\in\{1,...,s'\}$.
    Finally, $f_{\text{MCIP}}(\{\pmb{x}_i\},\pmb{y})$ can be recovered via computing $\frac{1}{2}\log(D)$.
\end{itemize}

\subsubsection{Correctness and Security}
Given the public parameter $\mathsf{pp}$, collected partially decrypted ciphertext $\{[ct]^{'}_{j}\}_{j\in\{1,..,s'\}}$, we have that
\begin{align*}
    D &= \frac{\prod_{i}ct_{i,0}^{\circ\pmb{y}_i}}{\prod_{i}\prod_{j}ct^{'}_{i,j,1} \cdot \prod_{j}(ct^{'}_{j,2})^2} \\
    % &= \frac{\prod_{i}(g^{\pmb{x}_{i}+\mathsf{H}(l)\pmb{U}_{i}} \circ (g^{\pmb{\alpha}^\intercal\pmb{W}_{i}})^{r_{i}})^{\circ\pmb{y}_i}}{\prod_{i}(ct_{i,1})^{\sum_{j}f^{(i)}(j)L_j(j)} \cdot (ct^{'}_{j,2})^{2\sum_{j}f^{(0)}(j)L_j(j)}} \\
    &= \frac{\prod_{i}(g^{\pmb{x}_{i}+\mathsf{H}(l)\pmb{U}_{i}} \circ (g^{\pmb{\alpha}^\intercal\pmb{W}_{i}r_{i}})^{\circ\pmb{y}_i}}{\prod_{i}(g^{r_{i}\pmb{\alpha}})^{f^{(i)}(0)} \cdot g^{2f^{(0)}(0)}}
    % &= \frac{\prod_{i\in[n]}g^{(\pmb{x}_{i} + \pmb{U}_{i})\circ\pmb{y}_i} \cdot \prod_{i\in[n]} (g^{\pmb{\alpha}^\intercal\pmb{W}_{i}r_{i}})^{\circ\pmb{y}_i}}{\prod_{i\in[n]}(g^{r_{i}\pmb{\alpha}})^{\langle\pmb{y}_i\pmb{W}_i\rangle} \cdot g^{2\sum^{n}_{i=1}\langle\pmb{y}_i\pmb{U}_i\rangle}} \\
    = \frac{\prod_{i}(g^{\pmb{x}_{i}+\mathsf{H}(l)\pmb{U}_{i}})^{\circ\pmb{y}_i}}{g^{2\mathsf{H}(l)\sum^{n}_{i=1}\langle\pmb{y}_i\pmb{U}_i\rangle}} \\
    &= \frac{g^{2\sum^{n}_{i=1}\langle\pmb{x}_i\pmb{y}_i\rangle} \cdot g^{2\mathsf{H}(l)\sum^{n}_{i=1}\langle\pmb{y}_i\pmb{U}_i\rangle}}{g^{2\mathsf{H}(l)\sum^{n}_{i=1}\langle\pmb{y}_i\pmb{U}_i\rangle}} 
    = g^{2f_{\text{MCFE}}(\{\pmb{x}_i\},\pmb{y})}\\
\end{align*}
For the proof of security of the threshold \textit{MCFE} scheme for $\mathcal{F}_{\text{MCIP}}$, as stated in Theorem~\ref{theo:tfe:miip}, we employ the same security definition as in \cite{abdalla2018multi}, namely, \textit{selective simulation-based security} (SEL-SIM security). We present the theorem and its proof in Section~\ref{sec:sp}.

% \vspace{-3mm}
\section{\textit{TAPFed} Framework}
\label{sec:TAPFed}

\subsection{Overview of \textit{TAPFed}}

\begin{figure}
    \centering
    \includegraphics[width=.45\textwidth]{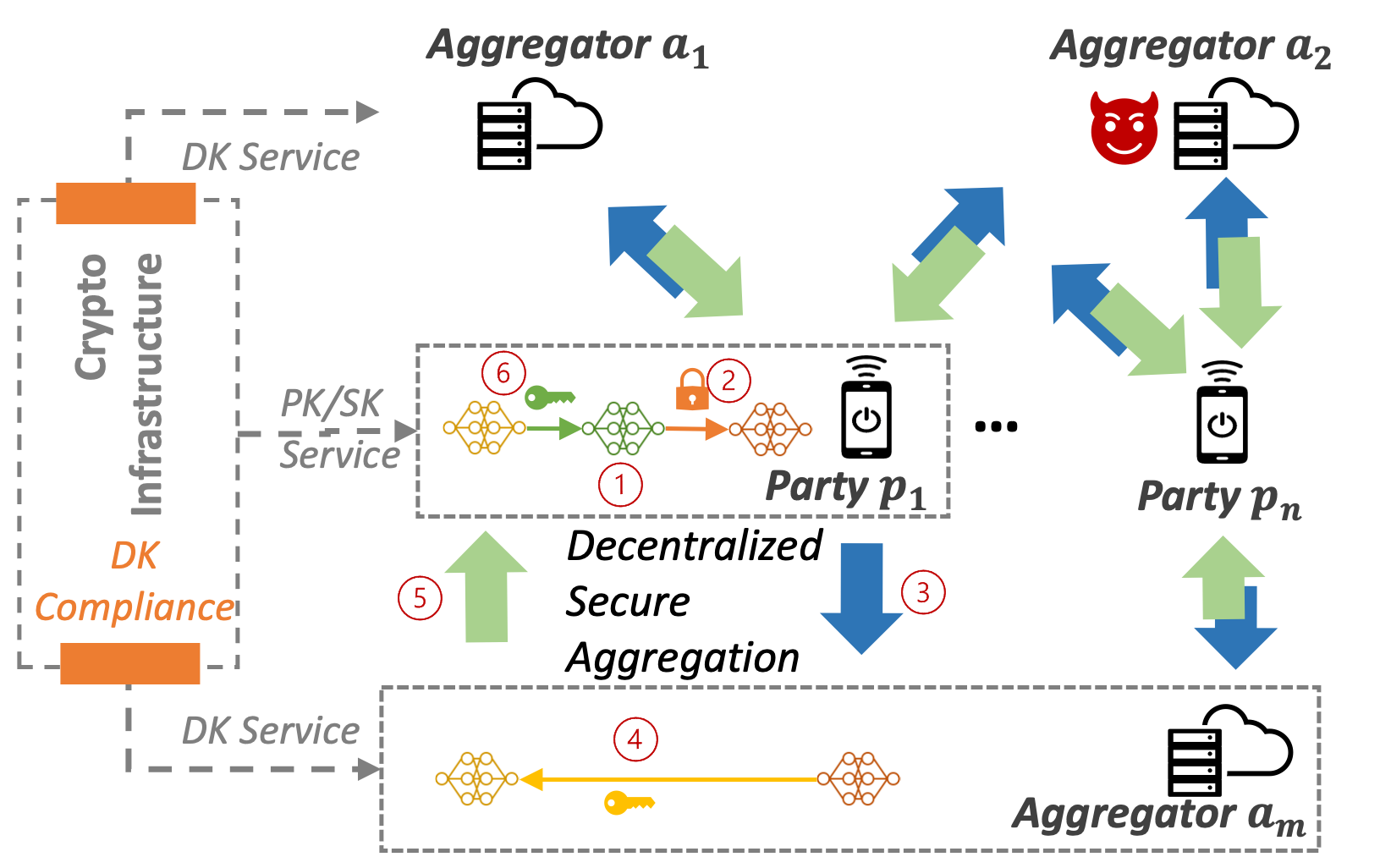}
    \vspace{-2mm}
    \caption{An overview of threshold secure aggregation in \textit{TAPFed} FL system. Note that there is NO NEED for aggregators to communicate with one another.}
    \label{fig:de-sec-agg}
    \vspace{-3mm}
\end{figure}

\noindent\textbf{How does \textit{TAPFed} work?}
\figurename~\ref{fig:de-sec-agg} depicts a summary of the threshold secure aggregation procedure and illustrates the operation of the \textit{TAPFed} FL system, which consists of a set of parties, decentralized aggregators (allowing a limited number of malicious aggregators), and a crypto infrastructure.
\textit{TAPFed} adopts threshold MCFE to achieve privacy-preserving FL coupled with a decentralized multi-aggregator setting to the secure aggregation process.

\textit{TAPFed} initiates threshold secure aggregation prior to the start of training by initializing crypto infrastructure, configuring cryptographic keys for each entity, and determining the agreed-upon fusion weight and training label for each FL training round.
To avoid redundancy, the following presentation begins with the threshold secure aggregation mechanism in \textit{iter-avg} fusion method, where \textit{iter-avg} fusion means that the aggregated model is the average of each party's model for each training round, and then discusses in Section~\ref{sec:tsecagg:fm} how our proposed mechanism supports other commonly used fusion methods in the FL system.

At each train round, each party trains a local model, and encrypts the model update using the \textit{threshold MCFE} cryptosystem that has been proposed in Section~\ref{sec:tfe} with the current training round as the \textit{cryptographic label}, and transmits it to each aggregator. 
Next, each aggregator performs the secure aggregation over the received encrypted model updates and returns the aggregated model update to each party. 
Notably, unlike existing MIFE-based FL solutions \cite{xu2019hybridalpha,xu2022detrust}, which exposes the aggregated model update to the aggregator, here the aggregated model update cannot be learned by aggregators because it is still in the ciphertext.
After receiving a set of aggregated model updates with a size greater than the specified threshold, each party is able to recover the aggregated model update in plaintext and start another training round until the maximum number of training rounds has been reached.

Importantly, in contrast to other innovative decentralized aggregator designs, \textit{TAPFed} eliminates the need for secure enclave hardware support while offering excellent scalability as a result of the design's elimination of additional communication beyond that between parties and aggregators.
This design, which avoids peer-to-peer aggregator communication, is consistent with the FL paradigm's primary setting as shown in Fig~\ref{fig:de-sec-agg}.
Moreover, \textit{TAPFed} directly supports both \textit{average} and \textit{weighted} fusion methods, whereas the vast majority of secure aggregation methods only support the former. 
This is discussed in Section~\ref{sec:tsecagg:fm}.

Note that \textit{TAPFed} primarily focuses on cross-silo federated learning, aligning with existing secure aggregation research based on computational cryptographic solutions \cite{xu2019hybridalpha,zhang2020batchcrypt}.

\noindent\textbf{Threat Model and Assumption}.
The following threat model is considered in \textit{TAPFed} framework:
\begin{itemize}[leftmargin=*]
    \item Unlike most existing PPFL solutions \cite{cheng2021separation,zhang2021shufflefl,rathee2023elsa,bell2023acorn,ma2023flamingo,lycklama2023rofl} that assume aggregator(s) are \textit{honest-but-curious} and don't collude with other aggregators in a two-server setting, \textit{TAPFed} eases this assumption. It allows for a limited number of adversarial aggregators enrolled in collusion, which isn't supported by many current privacy-preserving federated learning solutions. These aggregators may also sporadically or intentionally drop out during the FL training.
    
    \item Similarly to existing crypto-based PPFL solutions that either rely on a trusted dealer to synchronize keys among parties or on a trusted authority to provide key services to all entities in the FL framework, \textit{TAPFed} is also built on a trusted crypto infrastructure that is responsible for configuring the underlying cryptosystem and delivering keys to all entities in the framework.
\end{itemize}

We assume all communications take place over secure channels, effectively thwarting eavesdropping attacks. Unlike secure federated learning solutions that address security threats like model backdoor, poisoning, and stealing attacks, this work solely concentrates on privacy leakage risks posed by honest-but-curious or even adversarial aggregators. This approach aligns with existing privacy-preserving federated learning research; however, it excludes security or robustness-related attacks from its scope.

\noindent\textbf{Notation}.
% We introduce the following notation in the remainder of the paper.
Assume that the FL framework is composed of $n$ parties $S_\mathcal{P}$ and $m$ aggregators $S_\mathcal{A}$.
We use the terms $p_i \in S_\mathcal{P}$ and $a_j \in S_\mathcal{A}$ to refer to the party and aggregator, respectively, with party $p_i$ retaining its own dataset $\mathcal{D}_{i}$.
Let $\mathcal{M}$ be the machine learning model to be trained. 
Specifically, $\mathcal{M}^{(t)}_{a_j}$ and $\mathcal{M}^{(t)}_{p_i}$ denote the global aggregated and local trained models, respectively, at the $t$-th round of FL training, whilst the double-struck bracket $\llbracket\mathcal{M}\rrbracket$ denotes the encrypted model $\mathcal{M}$.
In addition, let $\widetilde{\mathcal{M}}_{p_i}$ be the local model injected with differential privacy noises. 
We utilize the $t_p$ and $t_a$ to denote the minimum number of parties and aggregators required for FL training.
In the following, we present more details of the privacy-preserving training process.

% \vspace{-2mm}
\subsection{Privacy-Preserving FL Training Process}
\label{sec:tsecagg:training}

First, we provide an overview of the privacy-preserving FL training framework, and then we go into detail about threshold secure aggregation, the fundamental component of \textit{TAPFed}.

\noindent\textbf{Algorithm Overview}. 
Algorithm \ref{alg:framework} gives an overview of how
\textit{TAPFed} works.
The system starts with the initialization of cryptosystem infrastructure, hyperparameters including training hyperparameters and fusion weight, and agreed-upon cryptographic training labels and keys (Lines 1-2).
For each training round, let's say the $k$-th round without sacrificing generality.
Party $p_i$ first trains local model $\mathcal{M}^{(k)}_{p_i}$ with local dataset $\mathcal{D}_i$, hyperparameters $hp$, and model from last round $\mathcal{M}^{(k-1)}_{p_i}$ (Lines 4-6).
Next, $p_i$ protects the local model using the proposed threshold decentralized secure aggregation (TDSA) mechanism, which will be described in detail later, and sends protected model update to all aggregators $a_j\in S_\mathcal{A}$ (Lines 7-8). 
After collecting responses from all parties, each aggregator $a_j$ performs secure aggregation using the TDSA mechanism and returns the aggregated model update to each party $p_i$ (Lines 9-12). 
Each party $p_i$ then recovers the aggregated model update in plaintext $\mathcal{M}^{(k)}_{\mathcal{A}}$, updates the local model $\mathcal{M}^{(k)}_{p_i}$, and conducts another round of FL training until the maximum number of training rounds has been reached (Lines 13-18).

\begin{algorithm}[t]
    \SetAlgoLined
    \caption{\textit{TAPFed} Training}
    \label{alg:framework}
    
    \SetKwInput{KwInput}{Input}
    \SetKwInput{KwOutput}{Output}
    \SetKwProg{Fn}{function}{}{}
    \SetKwRepeat{Do}{do}{while}
    
    \small
    \KwInput{Party set $S_\mathcal{P}$, each party $p_i \in S_\mathcal{P}$ has its dataset $\mathcal{D}_{i}$. Aggregator set $S_\mathcal{A}$. Maximum training rounds $q$. Hyperparameters $hp$. Encryption label for each traingin round $\{l^{(k)}\}_{k\in\{1,...,q\}}$.}
    \KwOutput{Final trained global model $\mathcal{M}_{G}$.}
    
    Initialize crypto-infrastructure\;
    Initialize hyperparameters $hp$, agreed-upon training labels $\{l^{(k)}\}_{k\in\{1,...,q\}}$, and cryptographic keys\;

    \Do{$k \le $ maximum training rounds $q$}{
        \ForEach{$p_i \in S_{\mathcal{P}}$}{
            \lIf{$k==1$}{$p_i$ initializes local model $\mathcal{M}^{(0)}_{p_i}$}
            $\mathcal{M}^{(k)}_{p_i} \gets$ trains model with $(\mathcal{D}_i, \mathcal{M}^{(k-1)}_{p_i}, hp)$\;
            % $\llbracket\mathcal{M}^{(k)}_{p_i}\rrbracket \gets$ $p_i$ performs \textit{tSecAgg-Protect}$(\mathcal{M}^{(k)}_{p_i})$\;
            $\mathtt{msg}^{(p_i)} \gets$ $p_i$ performs \textit{TDSA-Protect}$(\mathcal{M}^{(k)}_{p_i}, l^{(k)}$)\;
            $p_i$ sends $\mathtt{msg}^{(p_i)}$ to $a_j \in S_\mathcal{A}$\;
        }
        \ForEach{$a_j \in S_{\mathcal{A}}$}{
            $S^{(a_j)}_{\mathtt{msg}} \gets a_j$ collects responses from $p_i \in S_{\mathcal{P}}$\;
            $\mathtt{msg}^{(a_j)} \gets$ performs \textit{TDSA-Aggregate}$(S^{(a_j)}_{\mathtt{msg}}, l^{(k)})$\;
            $a_j$ sends $\mathtt{msg}^{(a_j)}$ to $p_i \in S_\mathcal{P}$\;
        }
        \ForEach{$p_i \in S_{\mathcal{P}}$}{
            $S^{(p_i)}_{\mathtt{msg}} \gets p_i$ collects responses from $a_j \in S_{\mathcal{A}}$\;
            $\mathcal{M}^{(k)}_{\mathcal{A}} \gets$ performs \textit{TDSA-Recover}$(S^{(p_i)}_{\mathtt{msg}})$\;
            $p_i$ updates $\mathcal{M}^{(k)}_{p_i} \gets \mathcal{M}^{(k)}_{\mathcal{A}}$\;
            \lIf{k==l}{
                 $\mathcal{M}_{G} \gets \mathcal{M}^{(k)}_{\mathcal{A}}$
            }
        }
    }
    \Return final model $\mathcal{M}_{G}$\;
\end{algorithm}

\noindent\textbf{Threshold Decentralized Secure Aggregation (TDSA).}
Algorithm \ref{alg:dsa} explains how \textit{threshold decentralized secure aggregation (TDSA)} mechanism operates, which is summarized as three functions: \textit{TDSA-Protect} and \textit{TDSA-Recover}, which are executed by parties; and \textit{TDSA-Aggregate}, which is carried out by aggregators.

In \textit{TDSA-Protect} function, each party employs threshold MCFE to encrypt local model update with its secret key $\mathsf{sk}_{i}$. 
Similar to the setting in \cite{xu2019hybridalpha}, TDSA is open and receptive to integration with the differential privacy (DP) mechanism in order to strengthen privacy guarantees (Lines 1-5).
In contrast to existing multi-input FE-based privacy-preserving solutions \cite{xu2019hybridalpha, xu2019cryptonn, xu2021nn}, our proposed threshold FE cryptosystem inherits from a multi-client FE design that incorporates cryptographic labels $\{l^{(k)}\}$ to prevent the cross-use of encrypted model updates from different training rounds.
Consequently, this design can thwart a variety of inference attacks and will be explained and analyzed in Section~\ref{sec:sp}.

In the \textit{TDSA-Aggregate} function, each aggregator first generates fusion weight $f_{a_j}$ based on the status of received model updates (Line 7). 
In the case of the \textit{iter-avg} fusion method without dropout consideration, for instance, $f_{a_j}$ is generated as an all-$\frac{1}{n}$ vector of size $n$, where $n$ is the number of parties.
Later in this section, we will explain how TDSA works in the case of dynamic parties and aggregators as well as other complex fusion methods.
Afterward, each aggregator $a_j$ requests the functional decryption key $\mathsf{dk}_{f}$ using $f_{a_j}$ and current training label $l^{(k)}$, while the crypto infrastructure $\mathcal{C}$ collects the key request materials and verifies their compliance (Lines 8-10).
If the verification is successful, $\mathcal{C}$ generates $\mathsf{dk}_{f}$ for aggregator $a_j$, and $a_j$ securely computes the aggregated model using the \textit{ShareDec} algorithm of the threshold MCFE scheme; otherwise, the function returns nothing (Lines 11-13).

The \textit{TDSA-Recover} function is responsible for recovering the aggregated model updates in plaintext by applying the \textit{CombineDec} algorithm of the threshold MCFE scheme with the securely computed model updates from $S^{'}_{a_j}$ (Lines 14-15).
Note that $S^{'}_{a_j}$ is merely a subset of the authored aggregator set $S_{a_j}$.

\begin{algorithm}[t]
    \SetAlgoLined
    \caption{Threshold Decentralized SA}
    \label{alg:dsa}
    
    \SetKwInput{KwInput}{Input}
    \SetKwInput{KwOutput}{Output}
    \SetKwProg{Fn}{function}{}{}
    \SetKwRepeat{Do}{do}{while}
    
    \small
    \KwInput{Crypto infrastructure $\mathcal{C}$ for threshold MCFE $\mathcal{E}$, public parameter $\mathsf{pp}$, $p_i\in S_{\mathcal{P}}$ is initialized with secret key $\mathsf{sk}_{i}$. Local training model $\mathcal{M}^{(k)}_{p_i}$ of party $p_i$ and encryption label $l^{(k)}$ at the $k$-th training round.}
    % \KwOutput{TODO}
    
    \Fn{TDSA-Protect($\mathcal{M}^{(k)}_{p_i}, l^{(k)}$)}{
        \lIf{DP not applied}{\Return $\mathcal{E}.\mathtt{Enc}(\mathsf{sk}_{i}, \mathcal{M}^{(k)}_{p_i}, l^{(k)})$}
        \Else{
            generate DP noise $\mathcal{N}^{(k)} \gets N^{\text{DP}}(\epsilon, \mathcal{M}^{(k)}_{p_i})$\;
            \Return $\mathcal{E}.\mathtt{Enc}(\mathsf{sk}_{i}, \mathcal{M}^{(k)}_{p_i}+\frac{\mathcal{N}^{(k)}}{n}, l^{(k)})$\;
        }
    }
    \Fn{TDSA-Aggregate($S^{(a_j)}_{\mathtt{msg}}, l^{(k)}$)}{
        $a_j$ prepares fusion weight $f_{a_j}$ based on $S^{(a_j)}_{\mathtt{msg}}$\;
        $a_j$ requests $\textsf{dk}_{f_{a_j}}$ from $\mathcal{C}$ with $(f_{a_j}, l^{(k)})$\;
        $\mathcal{C}$ collects $S_{(f,l)} \gets \{(f_{a_j}, l^{(k)})\}_{a_j \in S_{a_j}}$\;
        \If{$(f_{a_j}, l^{(k)})$ is complied with $S_{(f,l)}\setminus (f_{a_j}, l^{(k)})$}{
            $\mathcal{C}$ generates $\mathsf{dk}_{(f, l^{(k)})} \gets \mathcal{E}.\mathtt{DKGen}(\mathsf{msk}, f_{a_j}, l^{(k)})$\;
            \Return $\mathcal{M}^{(k)}_{a_j} \gets \mathcal{E}.\mathtt{ShareDec}(\mathsf{dk}_{(f, l^{(k)})}, S^{(a_j)}_{\mathtt{msg}}, f_{a_j})$\;
        }
        \lElse{ \Return None}
    }
    \Fn{TDSA-Recover($S^{(p_i)}_{\mathtt{msg}}$)}{
        \Return $\mathcal{M}^{(k)}_{\mathcal{A}} \gets \mathcal{E}.\mathtt{CombineDec}(S^{(p_i)}_{\mathtt{msg}})$\;
    }
\end{algorithm}

\noindent\textbf{Compliance of $\textsf{dk}_{(f,l)}$ Request}. 
To prevent potential abuse of functional decryption key (DK) service by part of curious aggregators, we have incorporated the \textit{DK Compliance} module within the crypto infrastructure $\mathcal{C}$.
Specifically, $\mathcal{C}$ first collects all DK requests, i.e., a set of tuples containing the fusion weight and training label, before processing them.
On the set of DK requests from aggregators, $\mathcal{C}$ identifies the majority of consistent DK requests $S^{'}_{(f,l)}$ whose size exceeds the specified trust threshold.
$\mathcal{C}$ then generates the functional decryption key $\mathsf{dk}_{(f,l)}$ based on the consistent fusion weight and training label for those aggregators whose DK requests have been complied with $S^{'}_{(f,l)}$. 
DK Compliance module can also inherit inference prevention features as illustrated in \cite{xu2019hybridalpha,xu2021nn}.

% \vspace{-3mm}
\subsection{Fusion Methods and Personalized FL}
\label{sec:tsecagg:fm}

\noindent\textbf{Supported Fusion Methods}. 
\textit{TAPFed} supports various fusion methods, such as commonly used weighted fusion method (e.g., FedAvg fusion method as shown in \cite{xu2019hybridalpha, xu2022detrust})  and average fusion method (e.g., \textit{IterAvg} fusion method adopted in \cite{truex2019hybrid, zhang2020batchcrypt, roth2021federated}).

As average fusion can be viewed as a special case of weighted fusion, only the latter will be explained here.
In the context of weighted average fusion, such as the FedAvg fusion method, one simple approach is to aggregate each party's local model update using the training sample size of each party as the fusion weight.
\textit{TAPFed} allows each party to transmit their training sample counts $s_{p_i}$ to every aggregator.
Then, each aggregator can compute the fusion weight vector as $\pmb{f}=(\frac{s_{p_1}}{\sum_i s_{p_i}}, ..., \frac{s_{p_n}}{\sum_i s_{p_i}})$ and request the crypto infrastructure for the functional decryption key $\mathsf{dk}_{\pmb{f}}$.
As previously explained, even though each aggregator can independently generate the fusion weight vector, the \textit{DK compliance} module will ensure the consistency of these vectors for each training round.
Note that we only discuss how to perform fusion weight generation in the static case; the dynamic case will be covered in the following Section~\ref{sec:tsecagg:dropout}.

\noindent\textbf{Personalized Secure Aggregation Support}.
Unlike conventional FL, which aims to train a single global model that may not always be preferable for all participating parties, personalized FL enables each party to only federate with other relevant parties to obtain a stronger model according to party-specific objectives.
In order to achieve this level of customization, it may be necessary for each party to train a local model with customized hyperparameters and for the aggregator to fuse model updates with party-specified fusion weights for each training round.

Our proposed decentralized secure aggregation does not interfere with the local training procedure; consequently, it supports naturally personalized FL with individualized local training.
Regarding personalized global fusion procedure, \textit{TAPFed} allows each party to specify a personalized fusion weight for a particular training round and share it with each aggregator in order to perform personalized secure aggregation.
Suppose that party $p_i$ specifies its fusion weight vector $\pmb{f}^{(p_i, k)}$ and agrees to personalized fusion weight vectors $\{\pmb{f}^{(p_j, k)}\}$ with other relevant parties $\{p_j\}$ for $k$-th training round.
Next, each party $p_i$ first encrypts local model update using the cryptographic labels $(p_i, l^{(k)})$ for its secure aggregation procedure and then encrypts model update with $(p_j, l^{(k)})$ specified for each relevant party $p_j$.
Each aggregator will then perform secure aggregation for encrypted model updates with label $(p_i, l^{(k)})$ and label $(p_j, l^{(k)})$, respectively. 
The threshold multi-client functional encryption scheme's algorithms dictate that only encrypted model updates with the same label can be aggregated. 
Finally, each party receives personalized and encrypted global model update fragments from the aggregators. They then select the fragments with a specified label and decrypt them using the corresponding decryption key to acquire the personalized global model.

% \vspace{-3mm}
\subsection{Dropout in \textit{TAPFed}}
\label{sec:tsecagg:dropout}

As previously noted, we present the \textit{TAPFed} framework for stable parties and aggregator groups.
This section explains how \textit{TAPFed} operates in the occasion of dropout.

\noindent\textbf{Dropout of Parties}.
Our decentralized secure aggregation solution does not require peer-to-peer communication between parties, so it naturally supports party dropout.
In the occasion that one party drops out during a particular training round, each aggregator could remove its fusion weight when requesting a functional decryption key in order to perform secure aggregation. 
For instance, if party $p_n$ drops out, we can set the fusion weight to $(f_{p_1}, ..., f_{p_{n-1}})$ instead of $(f_{p_1}, ..., f_{p_{n}})$.

\noindent\textbf{Dropout of Aggregators}.
As \textit{TAPFed} is based on the design of multiple decentralized aggregators, we also take aggregator failure into account.
\textit{TAPFed} enables $t$ stragglers among $m$ aggregators, where $t$ is the threshold that defines the maximum size of possible stragglers across all $m$ aggregators and usually is typically set to $\lfloor\frac{m}{2}+1\rfloor \le t \le m-2$.
For any training round, we support two types of dropouts: 
(\romannumeral1) aggregators drop out prior to receiving encrypted model updates; 
and (\romannumeral2) aggregators drop out after receiving encrypted model updates but are unable to perform secure aggregation.
The first scenario indicates that the aggregator fails before the secure aggregation, which can be viewed as our decentralized secure aggregation in the context of $m-t$ aggregators and is therefore naturally supported.
In the latter case, the aggregator drops out of the secure aggregation procedure; however, the underlying threshold MCFE scheme ensures that even with only $t$ of $m$ securely aggregated model updates, the authorized party can successfully recover the final aggregated model.

% \vspace{-2mm}
\section{Security and Privacy Evaluation}
\label{sec:sp}

\subsection{Security of threshold MCFE Scheme}
For the security of \textit{threshold multi-client functional encryption (tMCFE)} scheme $\mathcal{E}^{\mathcal{F}_{\text{MCIP}}}_{\text{tMCFE}}$, we employ the same security definition as in \cite{abdalla2018multi}. Theorem~\ref{theo:tfe:miip} states the security guarantee provided by  $\mathcal{E}^{\mathcal{F}_{\text{MCIP}}}_{\text{tMCFE}}$:
\begin{theorem}
\label{theo:tfe:miip}
    Assume an adversary that corrupts up to $t$-$1$ participants from the beginning; then, under the DDH assumption, $\mathcal{E}^{\mathcal{F}_{\text{MCIP}}}_{\text{tMCFE}}$ achieves selective simulation-based security.
    % Furthermore, a non-authorized player is not able to acquire the functionality result.
\end{theorem}
% Here, we use security proof methodology as in \cite{abdalla2018multi}, namely, simulation-based proof, to prove Theorem~\ref{theo:tfe:miip}.
% Then, we analyze security related to functionality result.

To prove the security of \textit{tMCFE}, we consider the following two cases: 
(\romannumeral1) $\adv$ can break one player (i.e., a \textit{sharing decryptor} or a \textit{combining decryptor});
(\romannumeral2) $\adv$ can corrupt up to $t$-$1$ players, including two sub-cases: (\romannumeral2.a) one \textit{combining decryptor} with $t$-$2$ \textit{sharing decryptors}; (\romannumeral2.b) $t$-$1$ \textit{sharing decryptors}.
Note that the \textit{sharing decryptor} and the \textit{combining decryptor} denote the entities running the \textit{ShareDecrypt} and \textit{CombineDecrypt} algorithms, respectively, as illustrated in Definition~\ref{def:th-feip}.
Then, we analyze the security of the scheme from two perspectives: encrypted data and functional results.

\noindent\textbf{Security for Encrypted Data}.
For the security of the encrypted data, we have the security claim, as presented in the first part of Theorem~\ref{theo:tfe:miip}.
Specifically, under the DDH assumption, $\mathcal{E}^{\mathcal{F}_{\text{MCIP}}}_{\text{tMCFE}}$ achieves selective simulation-based security (SEL-SIM-security).
Below is the formal proof in detail.

\begin{proof}
For case (\romannumeral1), the security threat is the same as that for the case of an ordinary multi-input FE scheme. 
Hence, we adopt the same security definition and advantage of the adversary $\adv$, as illustrated in \ref{sec:pre:fe}, in the formal proof below.

To prove the \textit{SEL-SIM-security} of \textit{tMCFE} scheme $\mathcal{E}_{\text{tMCFE}}$ for $\mathcal{F}_{\text{MCIP}}$, we need to prove that for any adversary $\adv$, $\advantage{\text{SEL-SIM}}{\adv,\mathcal{E}_{\text{tMCFE}}}=0$.
First, for the setup and encryption steps, we define the following simulator algorithms:
$\mathtt{Setup}^{\text{SIM}} = \mathtt{Setup}^{ot},$
$\mathtt{Encrypt}(\sk_{\text{m}}^{\text{SIM}}) = \pmb{u}_i,$
and for the generation of functional private key, we have the following simulator:
$\mathtt{KeyDerive}^{\text{SIM}}(\sk_{\text{m}}^{\text{SIM}}, \pmb{y}, \mathsf{aux}) \to sk_{\pmb{y}} = z,$
where $z$ is set as $\sum_{i\in[n]}\langle\pmb{u}_i\pmb{y}_i\rangle - \mathsf{aux} \mod L$. 

Then, we have the following for $\pmb{u}$ and $\pmb{u}-\pmb{x}$: $\forall \pmb{x}_i \in \{\pmb{x}_i\}_{i\in[n]}$, the distributions $\{\pmb{u}_i \mod L\}_{i \in [n]}$ and $\{(\pmb{u}_i - \pmb{x}_i) \mod L\}_{i\in[n]}$ are identical, where $\pmb{x}_i \in \ZZ_{L}$ and $\pmb{u}_{i} \sample \ZZ_{L}$.
Note that symbol $\sample$ denotes that $\pmb{u}_{i}$ is randomly sampled from $\ZZ_{L}$, and the independence of $\pmb{x}_i$ from the $\pmb{u}_i$ is only true
in the selective security game.
Hence, we have the simulator to rewrite the experiment $\textbf{REAL}^{\mathcal{E}^{ot}_\text{FE}}_\text{SEL}(1^{\lambda}, \mathcal{B})$ and oracle $\mathcal{O}_{H(\cdot)}$ as follows:

\begin{minipage}{.2\textwidth}
  \procedure[codesize=\footnotesize]{$\textbf{REAL}^{\mathcal{E}_\text{FE}^{ot}}_\text{SEL}(1^{\lambda}, \mathcal{B})$}{%
    \{x_i\}_{i\in[n]} \gets \mathcal{B}(1^{\lambda}, \mathcal{F}) \\
    \forall i\in[n]: \\
    \;\;\;\; \pmb{u}_i \sample \ZZ_{L} \\
    \;\;\;\; \pmb{ct}_i \gets \pmb{u}_i\mod L \\
    \alpha \gets \mathcal{B}^{\mathcal{O}_{H(\cdot)}}(\{ct_i\}_{i\in[n]}) \\
    \textbf{Output}: \alpha \\[1mm][\hline]
  }
\end{minipage}% This must go next to `\end{minipage}`
\begin{minipage}{.25\textwidth}
  \procedure[codesize=\footnotesize]{Oracle $\mathcal{O}_{H(\cdot)}$}{%
    z \gets \sum_{i\in[n]}\langle\pmb{u}_i,\pmb{y}_i\rangle - \langle\pmb{x}_i,\pmb{y}_i\rangle \mod L \\
    \Return: z \\[1mm][\hline]
  }
\end{minipage}

Therefore, the constructed $\textbf{REAL}^{\mathcal{E}_\text{FE}^{ot}}_\text{SEL}(1^{\lambda}, \mathcal{B})$ is also identical to the experiment $\textbf{IDEAL}^{\mathcal{E}_\text{FE}^{ot}}_\text{SEL}(1^{\lambda}, \mathcal{B})$ when executed with our simulator algorithms.
We can observe that the constructed oracle $\mathcal{O}_{H(\cdot)}$ corresponds to the oracle $\mathcal{O}(\cdot)$ (see Section \ref{sec:pre:fe}) that returns $\mathtt{KeyGerive}^{\text{SIM}}(\sk_{\text{m}}^{\text{SIM}}, \pmb{y}, \{\langle\pmb{x}_i,\pmb{y}_i\rangle\}_{i\in[n]}$ for every queried $\pmb{y}$.
Thus, we get $\advantage{\text{SEL-SIM}}{\adv,\mathcal{E}_{\text{tMCFE}}}=0$.
Hence, the adversary $\adv$ does not have the advantage of breaking the encrypted ciphertext.

For case (\romannumeral2), the adversary $\adv$ still has no advantage for breaking the encrypted ciphertext because $\adv$ in the above-illustrated simulation game is able to issue as many queries as expected; that is, the increase in the number of the corrupted players does not change such a situation. 
\end{proof}

\noindent\textbf{Security for Functionality Result}.
Next, we analyze the security related to the functionality result.
The \textit{tMCFE} scheme can ensure that the non-authorized player is not able to acquire the functionality result. 
Here, we do not consider the case (\romannumeral2.a) because $\mathcal{A}$ is not assumed to break the authorized entity that executes the combining decryption operation.

For case (\romannumeral2.b), suppose that $\mathcal{A}$ who corrupts $t$-$1$ players has non-negligible advantage $\epsilon$ to break the \textit{tMCFE} to acquire the functionality result.
In particular, the ``master'' functional private key is split into $s$ shares via Shamir's secret share scheme \cite{shamir1979share} in the TFE scheme, so that we can construct a simulator to transfer $\mathcal{A}$'s advantage to solve the $t$-\textit{of}-$s$ Shamir's secret share scheme with $t$-$1$ shares.
As proved in \cite{shamir1979share}, no adversary has a non-negligible advantage to solve that, and hence, $\mathcal{A}$ also does not have the non-negligible advantage to acquire the functionality result.

% \vspace{-3mm}
\subsection{Privacy Analysis of \textit{TAPFed}}
\label{sec:sp:privacy}

\noindent\textbf{Prevent Privacy Disclosure from Adversarial Aggregators}.
Attacks on federated learning systems typically aim to disrupt the model's functionality for security-related issues, while privacy-related attacks attempt to reveal personal data.
In this paper, we concentrate on privacy-related inference attacks and explore how TAPFed can thwart them. Specifically, we initially examine direct inference attacks like gradient and disaggregation assaults before discussing indirect attacks that could lead to privacy breaches, such as isolation, replay, and collusion attacks.

In the threat model, we consider a series of inference attacks (i.e., gradient inference,  disaggregation, isolation, replay and collusion attacks as demonstrated in recent privacy-preserving federated learning literature) caused by \textit{semi-trusted} aggregators $\mathcal{A}$.

\noindent\textit{(\romannumeral1) Gradient Inference Attack}: 
$\mathcal{A}$ launching a gradient inference attack against the FL system must acquire the model update in plaintext in order to conduct an inference analysis in a white/black box situation.
However, model updates from each party are protected by the \textit{tMCFE} cryptosystem, and as previously analyzed, the security of the proposed \textit{tMCFE} scheme ensures that no aggregator can acquire the aggregated model by independently breaching the cryptosystem. 
Therefore, the secure aggregation process of \textit{TAPFed} can prevent gradient inference attacks by design.

\noindent\textit{(\romannumeral2) Disaggregation Attack:} 
In the case of multi-round secure aggregation procedures, $\mathcal{A}$ may launch a disaggregation attack by solving a specific model update of the target party from multiple aggregation equations and then conduct a gradient inference attack as discussed above, even though the model updates are protected at each round.
As a prerequisite for a disaggregation attack against a secure aggregation-based PPFL, $\mathcal{A}$ must be able to acquire the intermediate aggregated model update, i.e., the aggregation result for each aggregation round.
As previously analyzed, each individual $\mathcal{A}$ is unable to acquire the intermediate aggregated model update, which is guaranteed by the underlying tMCFE cryptosystem, hence preventing disaggregation attacks.
Later in this section, the case of numerous aggregators' collusion will be discussed in detail.

\noindent\textit{(\romannumeral3) Isolation Attack:}
To emphasize this attack, consider $n=3$ parties.
At the $k$-th round of FL training, $\mathcal{A}_{j}$ may provide a malicious participation vector $(f_{\mathcal{A}_{j}}=(1,1,0), l^{(k)})$ for getting the functional decryption key $\mathtt{dk}_{f_{\mathcal{A}_{j}}}$ to isolate the third party's model update.
As stated in Algorithm~\ref{alg:dsa}, the trusted crypto infrastructure of \textit{TAPFed} will check the consistency of received participation vectors from all aggregators, which will detect and prohibit the aggregator from acquiring the function key $\mathtt{dk}_{f_{\mathcal{A}_{j}}}$ associated with the malicious participation vector $(f_{\mathcal{A}_{j}}, l^{(k)})$.
Without $\mathtt{dk}_{f_{\mathcal{A}_{j}}}$, $\mathcal{A}_{j}$ cannot conduct a secure aggregation.
Hence, $\mathcal{A}_{j}$ is incapable of launching the isolation attack.

\noindent\textit{(\romannumeral4) Replay Attack:}
As demonstrated in \cite{xu2022detrust}, by replaying the encrypted model update from various PPFL training rounds, $\mathcal{A}$ may potentially infer unexpected information.
For instance, $\mathcal{A}$ may feed encrypted model updates obtained from the $t$-th round into the ($t$+1)-th round secure aggregation procedure in order to infer additional knowledge.

\textit{TAPFed} can prevent replay attacks in the following ways: 
(a) \textit{TAPFed} does not reveal any intermediate aggregated model update to any individual aggregator, thereby preventing the demonstrated inference approach as demonstrated in \cite{xu2022detrust};
(b) \textit{TAPFed} utilizes the \textit{tMCFE} as its underlying cryptosystem, inheriting the essential characteristic of \textit{MCFE} schemes: model updates are encrypted with a label (e.g., round tag $l^{(k)}$ as shown in Algorithm~\ref{alg:dsa}).
By design, encrypted model update $\mathtt{Enc}(\mathcal{M}_{p_i}, l^{(t)})$ cannot be mixed with encrypted model update $\mathtt{Enc}(\mathcal{M}_{p_j}, l^{(t+1)})$ to perform partial decryption with $\mathtt{dk}_{f}$.
Accordingly, $\mathcal{A}$ cannot use encrypted model updates from previous rounds in any following round, preventing inference via replay attack.

\noindent\textit{(\romannumeral5) Collusion Attack:}
In contrast to prior analyses of a single adversarial aggregator, here we further examine the possibility that a limited number of aggregators may collude to infer private information.

Inheriting the threshold property of the proposed $t$-of-$n$ \textit{tMCFE} scheme, \textit{TAPFed} can thwart a collusion attack by a maximum of $t-1$ adversarial aggregators out of a total of $n$ aggregators in the FL training.
According to Section~\ref{sec:tfe}, a party can acquire the aggregated model update if and only if it obtains $t$ pieces of intermediate aggregated model updates from each aggregator.
In reverse, the intermediate aggregated model update cannot be disclosed to any $\mathcal{A}$ in the absence of at least $t$ colluding $\mathcal{A}$.
Hence, $\mathcal{A}$ cannot conduct the previously stated inference attacks without the intermediate aggregated model update.

\noindent\textbf{Prevent Privacy Leakage from Final Model}.
The primary objective of \textit{TAPFed} is to prevent privacy leakage during FL training; however, it can also provide a privacy guarantee to the final model. 
As explained in Section~\ref{sec:tsecagg:training}, \textit{TAPFed} can be simply incorporated with any differential privacy mechanism by injecting differential private noise into model update prior to secure aggregation.
Hence, \textit{TAPFed} can defend against inference attacks that target trained models within the FL framework with the privacy guarantee from the incorporated differential privacy mechanism.

% \vspace{-3mm}
\section{Experimental Evaluation}
\label{sec:eval}

% \vspace{-3mm}
\subsection{Experimental Setup}

\noindent\textit{\textbf{Implementation Consideration}}.
We used \textit{Python} programming language to implement the \textit{TAPFed}. The cryptographic components, including the proposed \textit{tMCFE} scheme, baseline cryptosystems, and key server were also implemented in Python using the \textit{gmpy2} library. This library is a C-coded Python extension module that supports multiple-precision arithmetic and relies on the GNU multiple precision arithmetic (GMP) library.
Furthermore, we utilize decryption optimization and floating-point number conversion strategies that are similar to those used in other functional encryption implementations \cite{xu2019hybridalpha,xu2021fedv,xu2022detrust}. 

\begin{table*}
    \centering
    \begin{threeparttable}
    \scriptsize
    \caption{Default Experimental Setting}
    \label{table:setting}
    \begin{tabular}{lllccccc}
        \toprule
            Experiments & Model Architecture & Parameters & Parties & FL Rounds & Local Epochs & Batch & Samples(train$\mid$test)/Party\\
        \midrule
            CNN-MNIST & 2xConv2D-MaxPooling-Flatten-2xDense & 1,199,882 & 5 & 20 & 10 & 50 & 5,000$\mid$2,000\\
            CNN-CIFAR10 & 2x(2xConv2D-MaxPooling)-Flatten-2xDense & 890,410 & 5 & 25 & 30 & 50 & 10,000$\mid$2,000\\
        \bottomrule
    \end{tabular}
    \vspace{-3mm}
    \end{threeparttable}
\end{table*}

\begin{figure*}[htb]
    \centering
    \includegraphics[width=\textwidth,trim=10 12 15 10, clip]{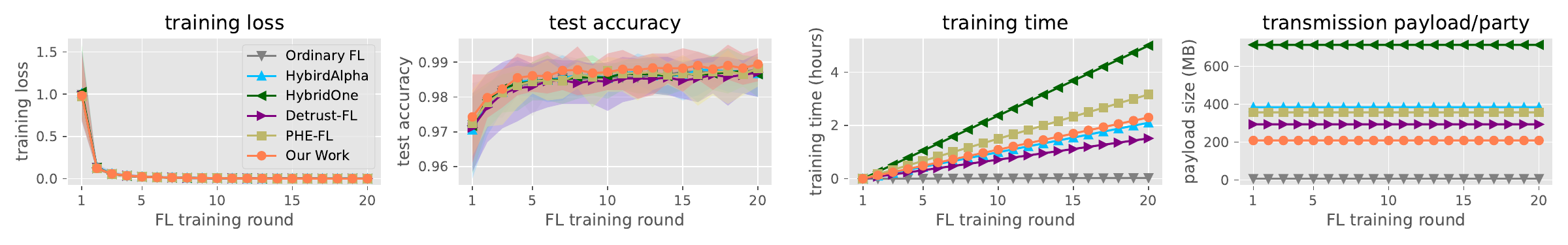}
    % \vspace{-8mm}
    \caption{Comparison of various baseline approaches in model training loss, model test accuracy, total training time and transmission payload on evaluating MNIST dataset. Note that the reported accuracy is the average accuracy of all parties and our \textit{TAPFed} employs 2 aggregators with the threshold set as 2.}
    \label{fig:cmp_baselines_mnist}
    \vspace{-3mm}
\end{figure*}

\begin{figure*}[htb]
    \centering
    \includegraphics[width=\textwidth,trim=10 12 15 10, clip]{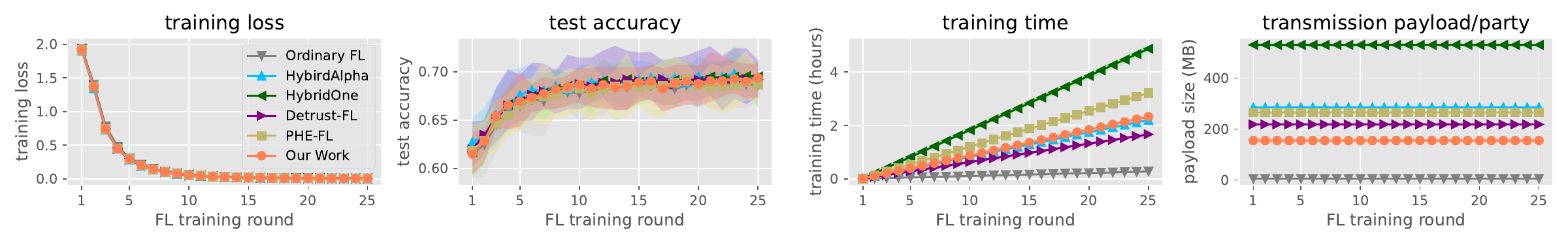}
    % \vspace{-8mm}
    \caption{Comparison of various baseline approaches in model training loss, model test accuracy, total training time and transmission payload on evaluating CIFAR10 dataset. Note that the reported accuracy is the average accuracy of all parties and our \textit{TAPFed} employs 2 aggregators with the threshold set as 2.}
    \label{fig:cmp_baselines_cifar10}
    \vspace{-3mm}
\end{figure*}

\noindent\textit{\textbf{Federated Learning Setting}}.
To evaluate the performance of the \textit{TAPFed} FL system, we train two varieties of Keras-based conventional neural network (CNN) models, with a total of 1,199,882 and 890,410 parameters, respectively, as described in \tablename~\ref{table:setting}, to classify the publicly available MNIST and CIFAR10 datasets.

In the configuration of the default FL setting, as partially described in \tablename~\ref{table:setting}, we set the following hyperparameters and settings: 
(i) For each experimental case, we set 5 parties and assigned dataset samples for each party in a non-iid setting. 
(ii) the total number of global FL training rounds is 20 for CNN-MNIST and 25 for CNN-CIFAR10 experimental cases;
(iii) for each global training round, we set 10 local epochs for CNN-MNIST and 30 local epochs for CNN-CIFAR10.
As the crypto-based secure aggregation operates on the \textit{integer} field, whereas model parameters are in \textit{floating-point} format, we have a comparable \textit{encoding precision} parameter, $pr$, to define the conversion scale factor between integers and floating-point numbers, namely, 
$\mathsf{encode}(i)=\lfloor i\times 10^{pr}\rceil$ and $\mathsf{decode}(i)=i/10^{pr}$. The default $pr$ is 4.

We compare our \textit{TAPFed} framework to various baselines:
\textit{(\romannumeral1)}: \textit{\textbf{Ordinary FL}} - training without any secure aggregation setting; 
\textit{(\romannumeral2)}: \textit{\textbf{PHE-FL}} \cite{zhang2020batchcrypt} - training using Paillier based secure aggregation setting.
\textit{(\romannumeral3)}: \textit{\textbf{HybridOne}} \cite{truex2019hybrid} - training using threshold Paillier based secure aggregation setting; 
\textit{(\romannumeral4)}: \textit{\textbf{HybridAlpha}} \cite{xu2019hybridalpha} - training using MIFE based secure aggregation setting;
\textit{(\romannumeral5)}: \textit{\textbf{DeTrust-FL}} \cite{xu2022detrust} - training using DMCFE based secure aggregation setting.
Note that, with the exception of our \textit{TAPFed}, the remaining baselines only work with a single aggregator setting.

% \vspace{-3mm}
\noindent{\textbf{\textit{Experimental Environment}}}.
We initially ran simulated tests on a local system equipped with an Intel(R) Core(TM) CPU i7-9700K, 8 cores, and 32GB of RAM, along with an NVIDIA GeForce RTX 2080 Ti GPU boasting 11G memory. It's important to note that network latency wasn't measured or reported in our simulated experiments since our framework operates within the same multi-process environment where each entity of our privacy-preserving federated learning framework is simulated by a separate process.
To further assess \textit{TAPFed}'s performance in a real distributed setting, we also carried out tests on the Aliyun cloud platform. Here we deployed 12 \textit{ecs.r6e.xlarge} instances equipped with Intel(R) Xeon(R) Platinum 8269CY CPU (4vCPU), and 32 GB of RAM.

\begin{figure*}[htb]
    \centering
    \includegraphics[width=\textwidth,trim=10 12 15 10, clip]{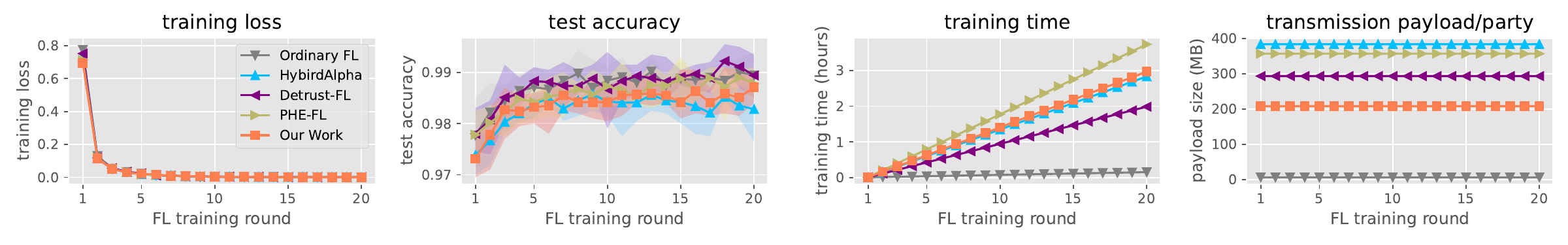}
    % \vspace{-5mm}
    \caption{Comparison of various baseline approaches in model training loss, model test accuracy, total training time and transmission payload on evaluating MNIST dataset in \textbf{\textit{real distributed setting}}. Note that the reported accuracy is the average accuracy of all parties.}
    \label{fig:cmp_baselines_mnist_dist}
    \vspace{-3mm}
\end{figure*}

\begin{figure*}[htb]
    \centering
    \includegraphics[width=\textwidth,trim=10 12 15 10, clip]{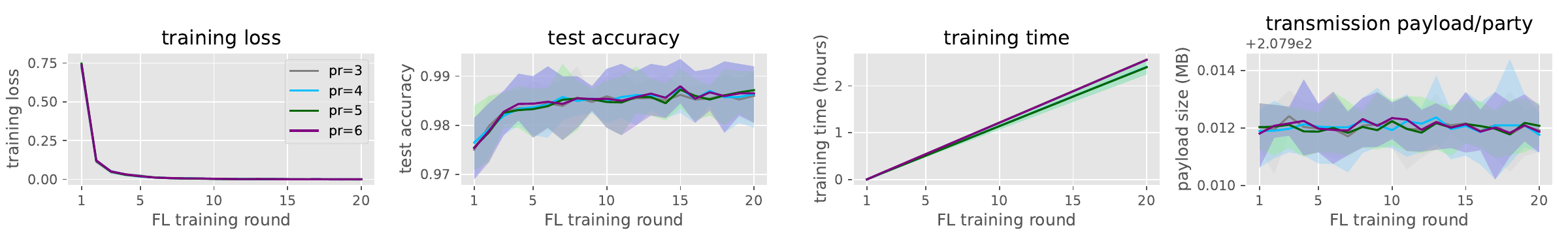}
    \vspace{-3mm}
    \caption{Impact of encoding precision of floating-point parameters in \textit{TAPFed} training and evaluation on MNIST dataset with the setting of 5 parties and 10 local training epochs per global FL training round.}
    \label{fig:cmp_precision}
    \vspace{-3mm}
\end{figure*}

\begin{figure*}[htb]
    \centering
    \includegraphics[width=\textwidth,trim=10 12 15 10, clip]{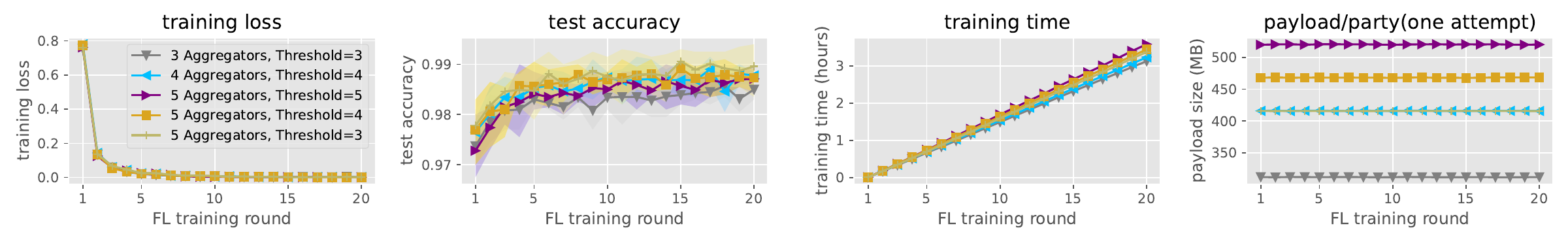}
    \vspace{-3mm}
    \caption{Impact of a number of aggregators in \textit{TAPFed} training and evaluation on MNIST dataset with setting of 5 parties and 10 local training epochs per global FL training round in \textit{\textbf{real distributed setting}}.}
    \label{fig:cmp_aggregators}
    \vspace{-3mm}
\end{figure*}

\begin{figure}[t]
    \centering
    \includegraphics[width=0.48\textwidth,trim=10 12 15 10, clip]{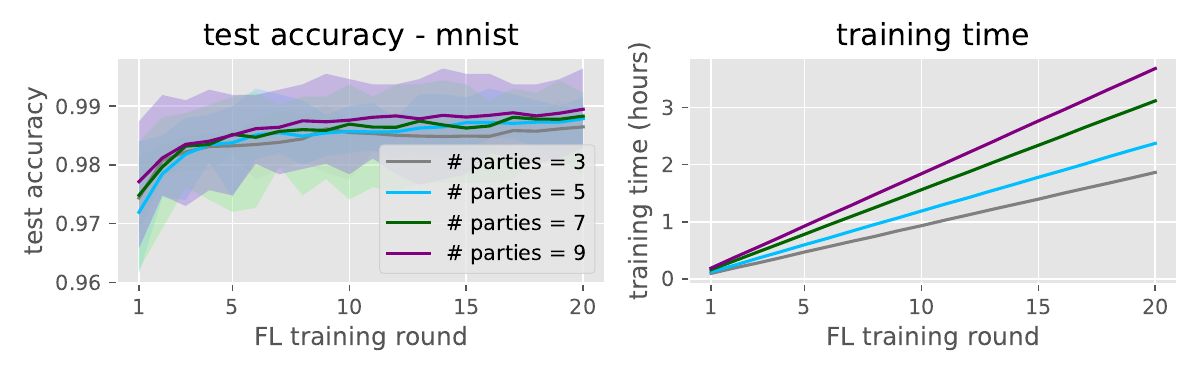}
    \vspace{-3mm}
    \caption{Impact of number of parties in \textit{TAPFed} training on MNIST dataset with settings of precision $=5$ and 5 local training epochs per training round.}
    \label{fig:cmp_parties}
    \vspace{-3mm}
\end{figure}

% \vspace{-3mm}
\subsection{Experimental Results}

\subsubsection{Performance Comparison to Baseline Approaches in Simulated Setting}

To evaluate \textit{TAPFed}'s performance relative to various baseline methods, we analyzed several metrics for each experimental case, including model training loss, test accuracy, training time, and transmission payload size. 
\figurename~\ref{fig:cmp_baselines_mnist} and \figurename~\ref{fig:cmp_baselines_cifar10} present the performance of two CNN models evaluating MNIST and CIFAR10 datasets, respectively. 
We conducted each experimental case at least three times and reported the minimum, maximum, and average results.

\noindent\textbf{Model Quality}. 
We report the model quality of \textit{TAPFed} with a comparison to five baselines using the MNIST and CIFAR10 datasets.
As shown in \figurename~\ref{fig:cmp_baselines_mnist}, we present the training loss and test accuracy on the MNIST dataset, over 20 global training rounds, where for each training round each party conducts 10 local training epochs. 
In terms of the CIFAR10 dataset of color images, as shown in \figurename~\ref{fig:cmp_baselines_cifar10}, we report the training loss and test accuracy over 25 global training rounds with 30 local training epochs.

According to the experimental results, we observe that
(i) The proposed threshold secure aggregation approach has no impact on the rate of learning convergence, training loss and test accuracy when compared to ordinary FL without any secure aggregation setting, as \textit{TAPFed} purely protects the exchanged model parameters without any noise introduction; 
and (ii) \textit{TAPFed} has achieved comparable model quality in comparison to other crypto-based secure aggregation baselines.

\noindent\textbf{Training Time}.
We also examine the impact of the secure aggregation approaches on training time, as shown in \figurename~\ref{fig:cmp_baselines_mnist} and \figurename~\ref{fig:cmp_baselines_cifar10} (third figure).
It shows that \textit{TAPFed} can achieve superior training time performance in comparison to (threshold) additive homomorphic encryption-based solutions and equivalent training time performance in comparison to \textit{HybridAlpha} \cite{xu2019hybridalpha}, which also utilizes functional encryption techniques. 
For example, in the context of the CNN-MNIST experimental case, \textit{TAPFed} reduces the training time by about 28\% and 55\% when compared to secure aggregation solutions of PHE-FL and HybridOne framework, respectively.

More critically, \textit{TAPFed} is capable of defending against newly emerging privacy inference attacks, such as isolation attacks, disaggregation attacks, etc.
Even though \textit{DeTrust-FL} has a relatively shorter training time, \textit{TAPFed}'s settings to prevent recently raised privacy inference attacks as illustrated in \cite{xu2022detrust} are simpler and more straightforward, whereas DeTrust\textit{-FL} relies on pre-configured participation matrix and large-scale enrolled parties.

\noindent\textbf{Transmission Overhead}.
To assess communication efficiency, we measure the transmission overhead during each training round between the aggregator and party. Specifically, we calculate the size of encrypted model updates sent by each party to the aggregator for every training round.

Out of all secure aggregation methods based on crypto techniques, \textit{TAPFed} has achieved the smallest transmission payload volume under the same security parameter configuration (i.e., 256) and model parameter size (i.e., 1,199,882 in CNN-MNIST model and 890, 410 in CNN-CIFAR10 model).
For instance, in the CNN-MNIST case, \textit{TAPFed} reduces transmission payload volume by 41.7\% and 70.8\%, respectively, compared to Paillier-based solutions such as \textit{PHE-FL} and \textit{HybridOne}. 
Additionally, when compared to functional encryption-based approaches like \textit{HybridAlpha} and \textit{DeTrust-FL}, \textit{TAPFed} decreases the volume of transmission payload by 45.8\% and 29\%, respectively.
Besides, we believe that the reason why \textit{TAPFed} achieved the smallest transmission payload volume but not the shortest training time is due to the fact that  \textit{TAPFed} introduces multiple aggregators, whereas our experiments were conducted in a single machine simulation setting as opposed to the real distributed setting.

\subsubsection{Performance Comparison to Baseline Approaches in Real Distributed Setting}

To further evaluate the performance of \textit{TAPFed} in a real distributed setting, we conducted experiments on the Aliyun cloud platform. 
Similar to the experiments in the simulated setting, we also compared \textit{TAPFed} with the various baselines in the distributed setting.
\figurename~\ref{fig:cmp_baselines_mnist_dist} presents how \textit{TAPFed} and different baselines performed on the MNIST dataset within an actual distributed scenario.

Compared to various baseline methods, the conclusions drawn from simulated settings also apply to distributed settings. 
To avoid repetition, we'll only highlight these observations in comparison with the simulated setting: (\romannumeral1) \textit{TAPFed}'s training time in a distributed setting is slightly longer than in a simulated one due to running on devices without GPUs, which may extend training time. Network latency and communication are also factors considered in our distributed setting; (\romannumeral2) The transmission payload volume of \textit{TAPFed} in the distributed setting closely matches that of its counterpart in the simulated environment, aligning with theoretical expectations.

\subsubsection{Performance of TAPFed}

In addition, we also examine the performance of \textit{TAPFed} in terms of federation impact, encoding precision and the number of parties.

\noindent\textbf{Impact of Encoding Precision}.
We conducted experiments with various encoding precision settings, ranging from $pre=3$ to $pre=6$, to investigate how encoding precision affects the conversion between FL training on floating-point formatted model parameters and secure aggregation on integer-based inputs.
Regarding training performance, \figurename~\ref{fig:cmp_precision} shows that there is no significant difference in terms of training loss and test accuracy among the various encoding precision settings. 
Additionally, the third and fourth figures in \figurename~\ref{fig:cmp_precision} indicate that different encoding precision settings do not have a significant impact on total training duration with respect to both training time and transmission payload.

\noindent\textbf{Impact of Number of Aggregatros}.
We further investigate the effect of aggregator size on \textit{TAPFed} performance by incorporating more aggregators into FL training than in previous experiments. Here, we maintain a constant party size of 5 and iterate the size of aggregators from 3 to 5 and its threshold setting, with identical precision settings and local training epoch sizes. As shown in \figurename~\ref{fig:cmp_aggregators}, augmenting the number of aggregators does not significantly affect model quality since secure aggregation functionality remains unaffected by aggregator size changes. We also note a minor increase in total training time as the number of aggregators rises due to an increased input requirement for local decryption of aggregated and encrypted model update fragments.
The communication overhead is linearly proportional to the number of aggregators, which is consistent with theoretical expectations.

\noindent\textbf{Impact of Number of Parties}.
We additionally examine the impact of party count on \textit{TAPFed} performance by enrolling a larger number of participants in FL training than in earlier experiments, with the same setting on precision and size of local training epochs.
As shown in \figurename~\ref{fig:cmp_parties}, increasing the number of parties can somewhat improve the model's precision, as more parties signify more data enrolled in the model's training, which results in a high-quality model.
As the training samples in our experiments were assigned randomly, the reported improvement in model accuracy is not as significant as in other FL research with imbalanced data settings.
Even though local model training and model encryption are carried out simultaneously by each party, as the number of parties grows, the total training time grows due to the increased number of inputs required for secure aggregation of encrypted model updates and the single-machine simulation setting across all experiments.

% \vspace{-3mm}
\section{Conclusion}
\label{sec:conclusion}
We have proposed \textit{TAPFed} to address the issue of privacy inference attacks raised by recent studies.
\textit{TAPFed} is constructed using our proposed threshold functional encryption techniques and is seamlessly integrated with existing FL platforms.
Our analysis demonstrates that \textit{TAPFed} can defend against recently proposed inference attacks caused by malicious aggregators to guarantee security and privacy. 
Our experimental study shows that \textit{TAPFed} offers equivalent performance in terms of test accuracy and training time while reducing the transmission overhead by 29\% compared to the state-of-the-art approaches.
In future work, we plan to explore other new cryptographic primitives to support more complex functions over encrypted data for supporting practical privacy-preserving applications.
Additionally, we aim to explore \textit{TAPFed}'s application in a cross-device federated learning environment.

\section*{Acknowledgment}
This work is funded by the National Natural Science Foundation of China, under grants No.62302022, No.62225202, No.62202038 and No.U22B2021. 
Additionally, Bo Li was supported by the Research on Soft and Hard Combinations of Multi-source Data Governance and MPC Algorithm (GJJ-23-007). 
The Beijing Advanced Innovation Center for Future Blockchain and Privacy Computing also partially supported the work.

\bibliographystyle{IEEEtran}
\bibliography{draft}

\end{document}